\documentclass{aa}

\usepackage{caption}
\usepackage{graphicx}
\usepackage{wrapfig}
\usepackage{float}
\usepackage{siunitx, physics}
\usepackage{tabularx}
\usepackage{soul}
\usepackage{dirtytalk}
\usepackage{geometry}
\usepackage{afterpage}
\usepackage{subcaption}
\usepackage{fancyhdr}
\usepackage{lscape}
\usepackage{setspace}
\usepackage{sidecap}
\usepackage{tablefootnote}
\usepackage{txfonts}

\newcommand{\NH}{\ensuremath{N_\textrm{H~I}}}
\newcommand{\NNa}{\ensuremath{N_\textrm{Na~I}}}
\newcommand{\NCa}{\ensuremath{N_\textrm{Ca~II}}}

\usepackage{hyperref}
\hypersetup{
    colorlinks=true,
    linkcolor=black,
    filecolor=black,      
    urlcolor=black,
    citecolor=black,
}

\graphicspath{ {./images/} }
\begin{document}

   \title{Probing neutral outflows in $z\sim2$ galaxies using JWST observations of \ion{Ca}{II} H and K absorption lines}

   \author{Caterina Liboni \inst{1},
          Sirio Belli \inst{1},
          Letizia Bugiani \inst{1,2},
          Rebecca Davies \inst{3,4},
          Minjung Park \inst{5},
          Charlie Conroy \inst{5},
          Razieh Emami \inst{5},
          Benjamin D. Johnson \inst{5},
          Amir H. Khoram\inst{1,2},
          Joel Leja \inst{6,7,8},
          Gabriel Maheson \inst{9,10},
          Matteo Sapori \inst{1,2},
          Trevor Mendel \inst{4,11},
          Sandro Tacchella \inst{9,10},
          Rainer Weinberger \inst{12}
          }

   \institute{Dipartimento di Fisica e Astronomia (DIFA), Università di Bologna,
              Bologna, Italy. \and 
              INAF, Osservatorio di Astrofisica e Scienza dello Spazio, Via Piero Gobetti 93/3, I-40129, Bologna, Italy. \and 
              Centre for Astrophysics and Supercomputing, Swinburne University of Technology, Hawthorn, Victoria, Australia. \and
               ARC Centre of Excellence for All Sky Astrophysics in 3 Dimensions (ASTRO 3D), Australia. \and
               Center for Astrophysics — Harvard \& Smithsonian, Cambridge, MA, USA. \and
               Department of Astronomy \& Astrophysics, The Pennsylvania State University, University Park, PA, USA. \and
               Institute for Gravitation and the Cosmos, The Pennsylvania State University, University Park, PA, USA. \and
               Institute for Computational \& Data Sciences, The Pennsylvania State University, University Park, PA, USA. \and
               Kavli Institute for Cosmology, University of Cambridge, Cambridge, UK. \and
               Cavendish Laboratory, University of Cambridge, Cambridge, UK. \and
               Research School of Astronomy and Astrophysics, Australian National University, Canberra, ACT, Australia. \and
               Leibniz Institute for Astrophysics, Potsdam, Germany.
             }

   \date{Received: June 06, 2025; Accepted: November 17, 2025}

  \abstract 
    {Using deep JWST/NIRSpec spectra from the Blue Jay survey, we performed the first systematic investigation of neutral gas content in massive galaxies at Cosmic Noon based on the Ca~II~H,~K absorption lines. We analyzed a sample of nine galaxies at $1.8 < z < 2.8$ with $\log M_\ast/M_\odot > 10.6$, for which we detected neutral gas absorption both in Ca~II and in Na~I. After removing the stellar continuum using the best-fit model obtained with \texttt{Prospector}, we fitted the excess absorption due to neutral gas in the Ca~II~H,~K doublet and in the Na~I~D doublet, together with nearby emission lines produced by ionized gas. We measured covering fractions between 0.2 and 0.9 from the Ca~II~H and K lines, which are spectrally well resolved in the NIRSpec $R\sim1000$ observations, unlike the absorption lines in the Na~I~D doublet. We measured the velocity shift, velocity dispersion, and column density separately for Ca~II and Na~I. About half of the galaxies present blueshifted Ca~II, indicative of an outflow of neutral gas and consistent with previous results based on Na~I. The velocity shift and the column density measured from Ca~II are correlated with those measured from Na~I, implying that these absorption lines trace gas in similar physical conditions. However, the column densities are not in a 1:1 relation, meaning that the relative amount of Ca~II and Na~I atoms along the line of sight varies with the gas column density. After investigating possible reasons for this behavior, we derived an empirical relation between the column density of Ca~II and the column density of Na~I and, in a more indirect way, of neutral hydrogen H~I. With this calibration, we measured the mass outflow rates in Blue Jay galaxies using the Ca~II lines, finding values in the range $\sim 2.7-56 \ M_{\odot}/yr$, broadly consistent (within a factor of roughly five) with previous studies of the Na~I lines. The employment of Ca~II lines offers a new way to infer properties of neutral gas from current and future JWST observations of massive galaxies at Cosmic Noon and beyond.
    }


   \authorrunning{Liboni et al.}
   \maketitle
    
\section{Introduction} \label{introduction}

Outflows are observed at all epochs, from the local Universe to galaxies at $z\sim6$ and beyond \citep[e.g.,][]{Shapley_2003,Veilleux_2005,Weiner_2009,Rubin_2014,Schreiber_2019}. They are typically driven by stellar feedback and/or active galactic nucleus (AGN) activity. Outflows have the capability to regulate the metal content of the interstellar medium (ISM) and, most significantly, to influence the process of star formation and quenching of galaxies. If the outflow velocity is sufficiently high to overcome the halo escape velocity, this can lead to the permanent removal of gas and the shutdown of star formation. Even when the velocity is insufficient to remove the gas from the halo, outflows may play a key role in the rapid quenching of star formation by moving cold gas from the inner parts of the galaxy to the hot circumgalactic medium \citep[e.g.,][]{Weinberger_2017,Man_Belli_2018,Trussler_2020,Zinger_2020}. 

Galaxies reach the peak of their star formation and feedback activity at $z\sim2$, the so-called Cosmic Noon \citep{Madau_2014}. Hence, Cosmic Noon is the best epoch to study outflows and their impact on galaxy evolution. 
Studies of emission lines, which trace gas in the warm ionized phase, have revealed that the mass outflow rate at $z~\sim~2$ is lower than what is required by theoretical models to suppress star formation \citep{Schreiber_2019,Lamperti_2021}. These measurements, however, do not include the contribution of cold gas, which is found in the molecular and neutral atomic phases \citep{Veilleux_2020}. In the local Universe, the neutral mass outflow rate is observed to be ten to one hundred times larger than the ionized one \citep{Roberts_Borsani_2020,Avery_2022}, suggesting that the ionized phase represents only a small fraction of the outflow mass budget. It is therefore essential to probe multiple gas phases when deriving the total mass outflow rate, but this is particularly challenging at high redshift. 

Recently, JWST observations have revealed the presence of widespread neutral outflows in high-redshift galaxies, particularly in massive recently quenched systems, where outflows are likely driven by AGN feedback \citep{Davies_2024, Belli_2024, deugenio_2024, Wu_2025, Valentino_2025}. 
These studies are based on the detection of resonant absorption lines due to neutral atomic gas, which were out of reach for previous ground-based observatories but have now been made accessible by the unprecedented sensitivity of JWST/NIRSpec.  
The measured neutral mass outflow rates are on the order of $\sim 10-100~M_{\odot}$/yr, substantially higher than what previous studies had found for the ionized phase. These new observations suggest that galaxy outflows may be sufficiently powerful to cause the quenching of star formation once the cold phase is taken into account. However, measurements of neutral outflows at Cosmic Noon are still affected by large systematic uncertainties due to an unknown covering fraction, outflow geometry, and gas abundances.

Many neutral outflow measurements are based on the study of blueshifted absorption in the Na~I~D doublet (Na~I $\lambda\lambda$5890, 5896~$\AA$), which has long been used to probe neutral outflows in the local Universe due to its strength and convenient location in the optical spectrum \citep{Heckman_2000_Na_blueshift,Rupke_2002_Na_blueshift,Rupke_2005_sample,Rupke_2005_analysis,Martin_2005,Roberts_Borsani_2019,Concas_2019}. Moreover, being an absorption line doublet, Na~I~D provides substantially more information compared to an individual transition, enabling one to break the degeneracy between covering fraction and column density.
However, the Na~I~D doublet is often heavily blended due to the small wavelength separation between the two absorption lines. This can considerably limit the accuracy of two measurements that are crucial for calculating the mass outflow rate: 1) the kinematics of the absorption lines; and 2) the equivalent widths of the two separate doublet components, which are in turn used to derive the column density and the covering fraction of the neutral gas. To overcome this limitation, it is necessary to detect other doublets with wider wavelength separations. A popular choice is Mg~II $\lambda\lambda$2796,2803$\AA$, which is in the rest-UV, and can thus be observed with optical instruments up to $z\sim2.5$ \citep{Tremonti_2007, Weiner_2009, Rubin_2010}. However, Mg~II and other rest-UV transitions (such as those due to Fe~II) require bright UV continuum emission by the source, which is not present if the galaxy is quiescent and/or dusty. Moreover, when multiple transitions are detected in the same galaxy, they can yield mass outflow rates that differ by more than an order of magnitude \citep{Rubin_2010, Valentino_2025}. This is due to both observational limitations (saturation effects, imprecise dust depletion correction) and physical reasons, since each species traces a different range of gas temperatures. If we want to understand the systematics involved in the measurement of neutral outflows, employing a wide range of absorption lines is clearly a priority.

In this work, we explore the use of the Ca~II H and K absorption lines (i.e., the Ca~II~$\lambda\lambda$3934, 3969$\AA$ doublet) as a tracer of outflows in massive galaxies at Cosmic Noon. This doublet probes neutral gas, since the ionization potential for Ca~II (11.9 eV) is lower than the hydrogen ionization potential. The main limitation of the Ca~II H and K lines is that they are also present in the stellar spectrum, and are very strong for old stellar populations. This makes it difficult to detect the excess absorption from neutral gas superimposed on the stellar absorption lines. For this reason, Ca~II H and K are typically not used in studies of local galaxies. This challenge is partially mitigated at high redshift, where stellar populations are young and have relatively weaker Ca~II absorption lines, although contamination is still present due to absorption and/or emission in the H$\epsilon$ line, which happens to be nearly coincident with the Ca~II~H wavelength. The advantage of using Ca~II H and K lines as diagnostics of neutral gas is that they are redward of the Balmer break, and can therefore be easily detected against a relatively strong continuum; moreover, their wide wavelength separation makes it easy to spectrally resolve the two lines.

The feasibility of detecting the neutral gas contribution to the Ca~II H and K lines has been demonstrated by \citet{Belli_2024} using JWST spectroscopy for COSMOS-11142, a massive galaxy at $z=2.45$. In that galaxy, the neutral gas kinematics derived from Ca~II H, K matches the Na~I D kinematics, thus confirming that the two doublets probe the same type of gas. In the present study, we extend the analysis of the Ca~II H and K lines and the comparison to the Na~I D properties to the full sample of massive galaxies in the Blue Jay survey, from which COSMOS-11142 was drawn.
The description of the survey and the criteria used for the sample selection are reported in Section~\ref{Data_Sample_selection}, while Section~\ref{abs_line_fitting} illustrates the method used for the analysis of Na and Ca absorption lines. The gas kinematics are analyzed in Section~\ref{kinematics} and the column densities in Section~\ref{column_dens_section}. A calibration of Ca~II H, K as a probe of the neutral gas mass is discussed in Section~\ref{discussion}, and a summary of this work is presented in Section~\ref{conlusions}.

\begin{figure}[t]
    \centering
    \includegraphics[width=\linewidth]{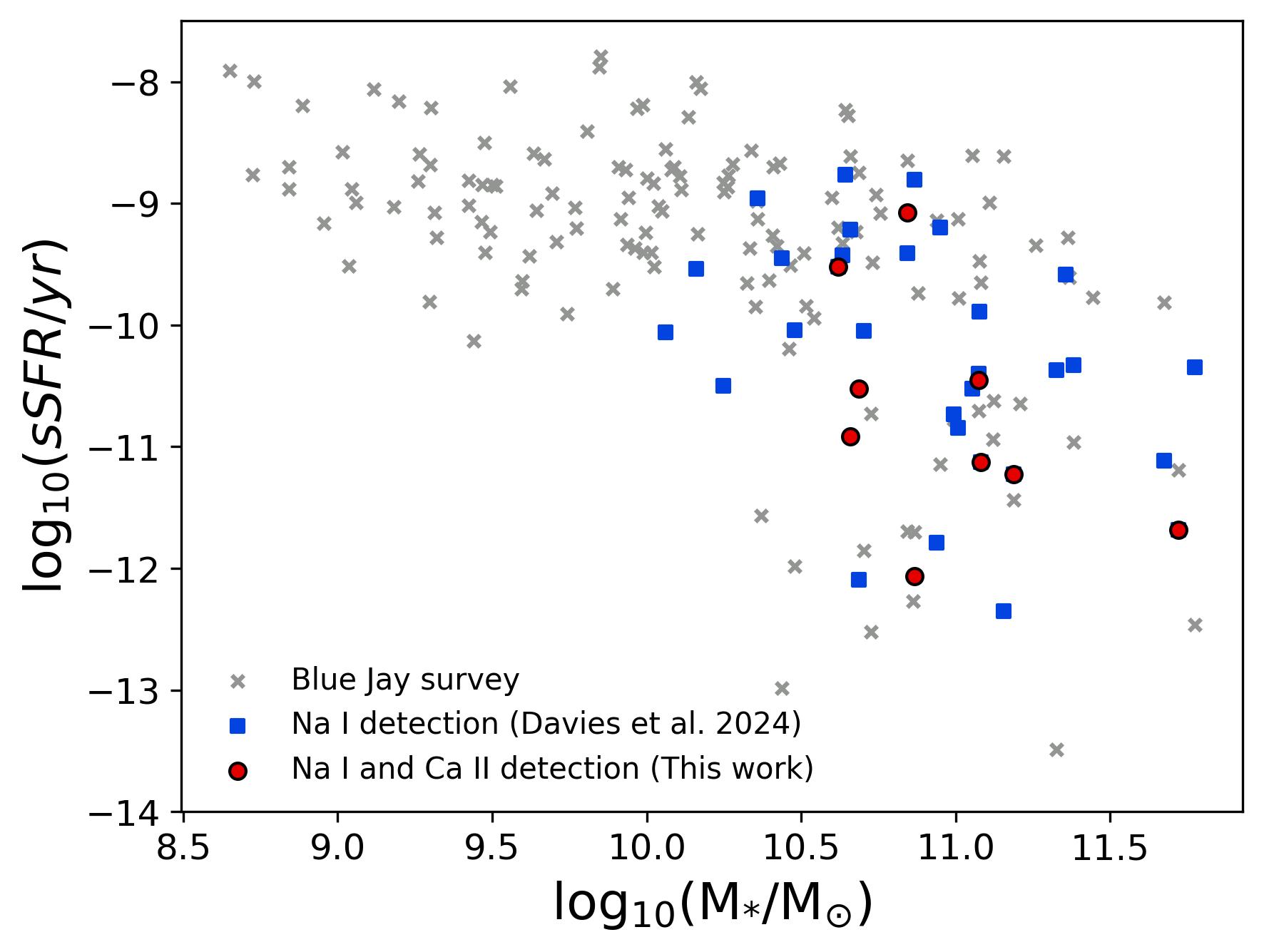}
    \caption{Distribution of stellar mass and sSFR for 141 galaxies from the Blue Jay survey. Gray crosses mark systems with no detection of excess Na~I~D absorption. Blue squares represent the sample of galaxies with excess Na~I~D absorption due to neutral gas studied by \citet{Davies_2024}. The galaxies analyzed in the present work are represented by red circles, and they have both Na~I and Ca~II absorption excess clearly detected.}
    \label{fig:sample}
\end{figure}

\begin{table*}
\renewcommand{\arraystretch}{1.7}
\caption{Sample properties from \texttt{Prospector} fit and \texttt{emcee} inference.}
\label{table_gal_values}
\centering
\begin{tabular}{ccccccccccc}
\hline \hline
IDs & $z$ & log(M$_{*}$/M$_{\odot}$) & sSFR & log(\NCa) & log(\NNa) & $\sigma_{Ca \ II}$ & $\sigma_{Na \ I}$ & $\Delta$ V$_{Ca \ II}$ & $\Delta$ V$_{Na \ I}$ & C$_{f}$\\
\hline
 \ & \ & \ & Gyr$^{-1}$ & \ & \ & km/s & km/s & km/s & km/s & \ \\
\hline
8002 & 2.7 & 10.7 & 2.4 & 13.7$^{+0.8}_{-0.4}$ & 13.9$^{+0.2}_{-0.1}$ & 359$^{+80}_{-101}$ & 219$^{+75}_{-79}$ & -180$^{+164}_{-158}$ & -451$^{+57}_{-36}$ & 0.4$^{+0.4}_{-0.2}$\\
18668 & 2.1 & 11.2 & 0.004 & 14.4$^{+0.3}_{-0.3}$ & 14.1$^{+0.2}_{-0.1}$ & $<137$\tablefootmark{a} & 348$^{+65}_{-88}$ & -69$^{+43}_{-50}$ & -393$^{+175}_{-83}$ & 0.6$^{+0.2}_{-0.1}$\\
11142 & 2.4 & 11.1 & 0.02 & 13.6$^{+0.3}_{-0.2}$ & 13.8$^{+0.1}_{-0.1}$ & 92$^{+28}_{-24}$ & 87$^{+28}_{-25}$ & -125$^{+26}_{-26}$ & -222$^{+26}_{-26}$ & 0.7$^{+0.2}_{-0.2}$\\
16874 & 2.1 & 10.6 & 0.6 & 14.0$^{+0.5}_{-0.3}$ & 13.3$^{+0.1}_{-0.2}$ & 115$^{+107}_{-55}$ & 182$^{+109}_{-66}$ & -126$^{+70}_{-70}$ & -141$^{+110}_{-91}$ & 0.5$^{+0.3}_{-0.2}$\\
9395 & 2.1 & 10.9 & 0.002 & 13.6$^{+0.5}_{-0.3}$ & 12.9$^{+0.2}_{-0.4}$  & 247$^{+97}_{-83}$ & 350$^{+97}_{-141}$ & 102$^{+89}_{-101}$ & -62$^{+203}_{-230}$ & 0.5$^{+0.3}_{-0.3}$\\
18071 & 2.8 & 10.7 & 1.8 & 13.7$^{+0.4}_{-0.2}$ & 13.1$^{+0.2}_{-0.5}$ & 194$^{+74}_{-53}$ & 285$^{+123}_{-127}$ & -122$^{+59}_{-62}$ & -234$^{+229}_{-171}$ & 0.6$^{+0.3}_{-0.3}$\\
16419 & 1.9 & 11.7 & 0.006 & 13.6$^{+0.6}_{-0.5}$ & 13.3$^{+0.1}_{-0.1}$ & 62$^{+121}_{-49}$ & 406$^{+57}_{-84}$ & -14$^{+87}_{-91}$ & -35$^{+108}_{-127}$ & 0.5$^{+0.3}_{-0.3}$\\
10245 & 1.8 & 10.8 & 2.3 & 14.0$^{+0.5}_{-0.3}$ & 13.5$^{+0.1}_{-0.1}$ & 291$^{+116}_{-101}$ & 113$^{+87}_{-40}$ & -19$^{+93}_{-88}$ & -102$^{+63}_{-66}$ & 0.5$^{+0.3}_{-0.2}$\\
19572 & 1.9 & 11.1 & 0.2 & 14.2$^{+0.4}_{-0.2}$ & 13.8$^{+0.1}_{-0.1}$ & 317$^{+97}_{-99}$ & 86$^{+33}_{-22}$ & 39$^{+72}_{-87}$ & 104$^{+21}_{-21}$ & 0.6$^{+0.3}_{-0.2}$\\
\hline
\end{tabular}
\tablefoot{The values of $z$, log(M$_{*}$/M$_{\odot}$) and sSFR are from the \texttt{Prospector} fit to the observed photometry and spectroscopy. The measures of log(\NCa), log(\NNa), $\Delta$ V$_{Ca \ II}$, $\Delta$ V$_{Na \ I}$, C$_{f}$ are from fitting our model of neutral and ionized gas to the normalized spectrum. Column density, velocity dispersion, and velocity shift are given separately for the fits to the Ca~II and Na~I doublets. The covering fraction C$_f$ is from the Ca~II fit, and is adopted as a fixed value in the Na~I fit. $\sigma_{Ca \ II}$ and $\sigma_{Na \ I}$ are the observed velocity dispersions corrected for the nominal spectral resolution, which is likely an underestimate because it assumes uniform slit illumination \citep{Degraaff_2024_resolution}.\\
\tablefoottext{a}{The observed velocity dispersion is smaller than the nominal one, so we adopted the observed value as an upper limit.}}
\end{table*}

\section{Data and Sample selection} 
\label{Data_Sample_selection}

\subsection{Blue Jay survey}
In this study, we analyze spectroscopic data from the Blue Jay survey, a JWST Cycle-1 program (GO 1810; PI Belli). The aim of this survey is to study the stellar populations and the ISM of 153 galaxies at Cosmic Noon. The spectra were acquired with the NIRSpec Micro-Shutter Assembly \citep[MSA,][]{Ferruit_2022}, adopting a spectral resolution R~$\approx 1000$. This was achieved using three medium resolution gratings (G140M, G235M, G395M) with exposure times of 13~h, 3.2~h, and 1.6~h, respectively. The entire wavelength coverage provided by the gratings is $1~-~5~\mu \rm m$, with some small gaps due to the physical separation between the two NIRSpec detectors. The observations were performed placing a slitlet with at least two MSA open shutters on each target and applying a two-point A-B nodding pattern along the slit. Empty shutters were used to provide a master background, which was then subtracted from each spectrum. The spectroscopic data reduction was performed with a modified version of the JWST Science Calibration Pipeline v1.10.1 \citep{Bushouse_2023}, using version 1093 of the Calibration Reference Data System.

The Blue Jay sample was selected in the COSMOS field using Hubble Space Telescope (HST) observations provided by CANDELS \citep{Grogin_2011,Koekemoer_2011}, together with the photometric catalog of the 3D-HST team \citep{Skelton_2014}, which includes ground- and space-based data. We exclude four filler targets at $z\sim6$ and 8 galaxies for which spectral extraction failed, resulting in a final sample of 141 galaxies with stellar masses 9~\textless~$\log(M_{*}/M_{\odot})$~\textless~11.5 and redshifts $1.7<z<3.5$. The sample is representative of the parent galaxy population because it is selected to have a roughly uniform coverage in mass and redshift, and is not biased in color, morphology, or specific star formation rate (sSFR). Additional details on the target selection, observations, and data reduction for the Blue Jay survey will be provided in \citet{Belli_survey}.

\begin{figure*}[htbp] 
\centering
\begin{minipage}[c]{0.6\textwidth} 
\centering 
\includegraphics[width=\linewidth]{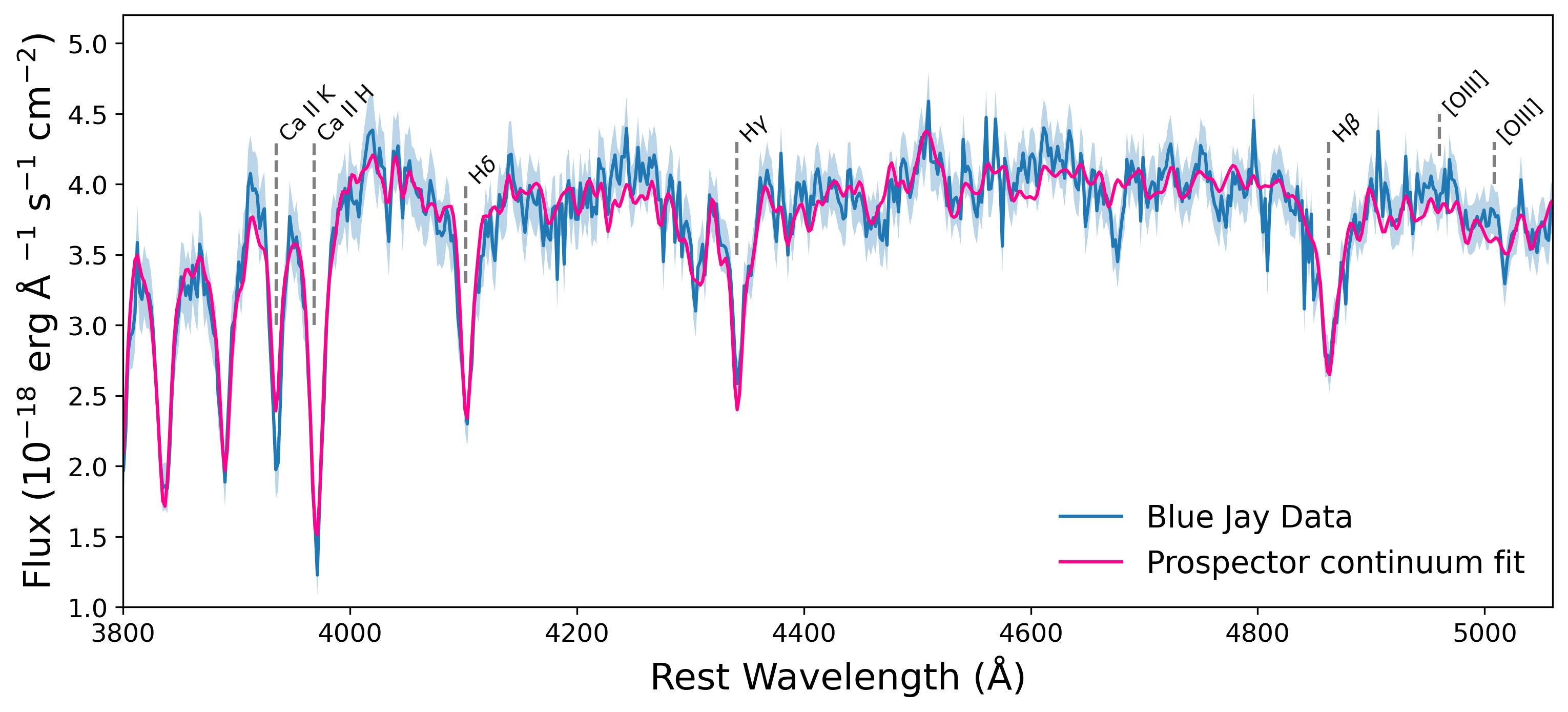}
\vspace{0.5cm} 
\includegraphics[width=\linewidth]{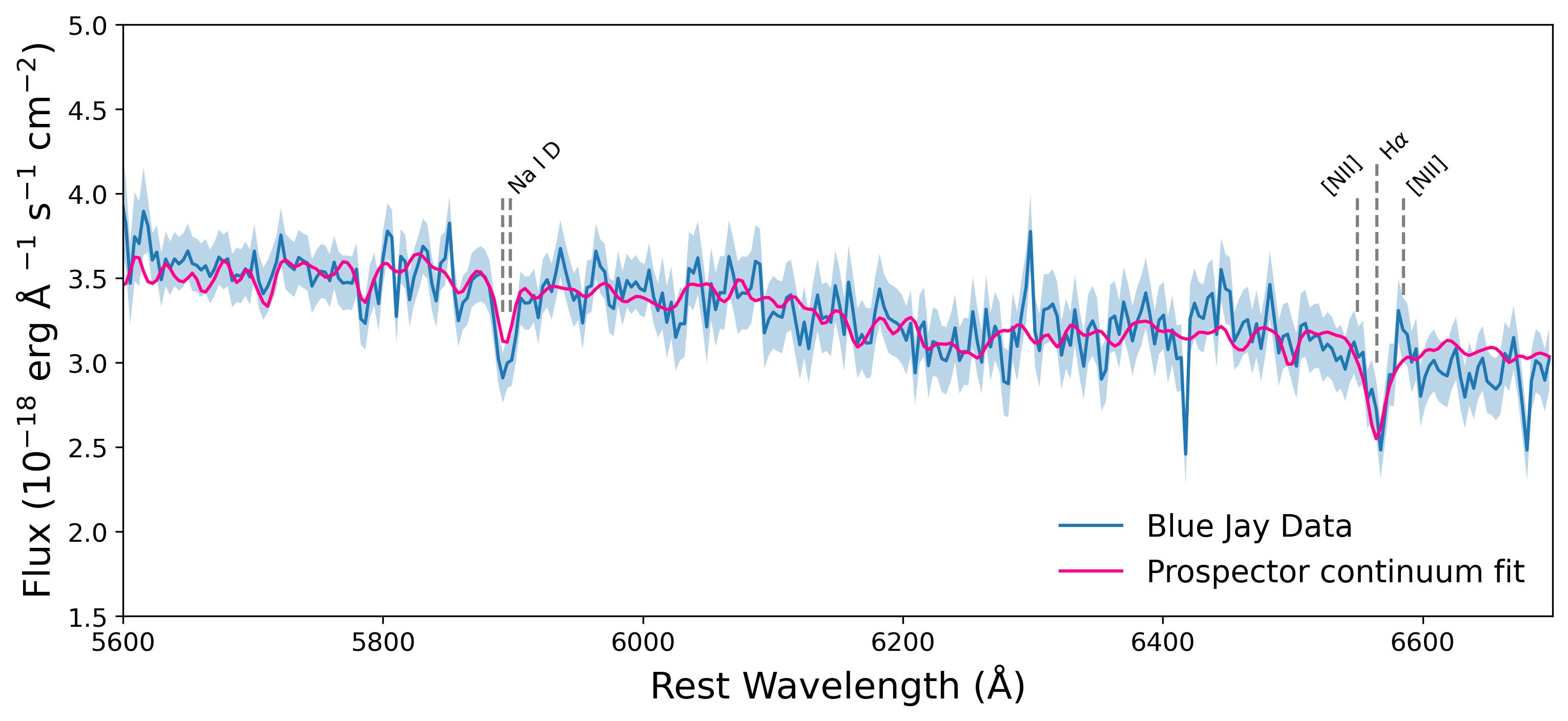} \end{minipage}
\hfill
\begin{minipage}[c]{0.35\textwidth} 
\caption{NIRSpec spectrum of galaxy COSMOS 9395 at $z \sim 2.1$. The observed spectrum is shown in blue, and the shaded area is the 1$\sigma$ uncertainty, while the dark pink curve is the best-fit stellar continuum provided by \texttt{Prospector}. Some of the most important emission and absorption lines visible in this wavelength range, including Ca~II~K, Ca~II~H and Na~I~D, are highlighted.} \label{fig:observed_spec_9395} 
\end{minipage} 
\end{figure*}

\subsection{Stellar population modeling}

The Na~I D and Ca~II H, K absorption lines observed in galaxy spectra are produced both by stars and by neutral gas in the ISM and/or the outflow. In order to study the neutral gas, it is thus first necessary to remove the stellar contribution to the observed absorption lines, which requires an accurate model of the stellar spectrum.
For each galaxy in the Blue Jay survey, we obtained a stellar population model (together with inferred stellar population and dust properties) using \texttt{Prospector} \citep{Johnson_2021}. We adopted the FSPS stellar population synthesis library \citep{Conroy_2009, Conroy_2010} with the MIST isochrones \citep{Choi_2016} and the C3K spectral library \citep{Cargile_2020}, a \citet{Chabrier_2003} initial mass function, a non-parametric star formation history with 14 age bins and a continuity prior \citep{Leja_2019}, dust attenuation and dust emission. The models are then fit to the observed NIRSpec spectroscopy together with space-based broadband photometry from HST and JWST/NIRCam \citep{Skelton_2014}. The Na~I~D and Ca~II~H,~K lines are masked in the spectrum, together with emission lines due to ionized gas. Due to their small size, the MSA shutters can only capture a fraction of the light emitted by the target. Therefore, during the \texttt{Prospector} fit a multiplicative polynomial correction is applied to the spectrum to match it with multi-band photometric data. This method simultaneously provides a flux calibration and a slit-loss correction.
An exhaustive description of the \texttt{Prospector} fits is given in \citet{Park_2024}; the only difference with that work is that here we consider a wider wavelength range when fitting the spectroscopy (3700~\AA\ to 13700~\AA\ in the rest frame), as done in \citet{Bugiani_2024}. The fit results include the best-fit model of the stellar spectrum across the full wavelength range of the NIRSpec observations, together with measurements of physical properties such as stellar mass and dust attenuation.

\subsection{Sample selection}
\label{sec:sample}

The goal of this study is to compare the neutral gas properties derived from Ca~II H and K to those derived from Na~I~D. For this reason, we start with the sample of 30 galaxies from the Blue Jay survey in which excess absorption in Na~I due to neutral gas was detected by \citet{Davies_2024}. We then applied a series of further cuts to obtain the final sample:
\begin{itemize}
    \item We removed two galaxies in which the Ca~II lines fall in a detector gap, and 2 galaxies with poor data reduction quality, preventing an accurate fitting of the absorption lines of interest.
    \item For some galaxies, the signal-to-noise ratio (S/N) in the Ca~II H and K wavelength region is too low to robustly subtract the stellar absorption lines and obtain meaningful constraints on the neutral gas properties. We set this threshold at S/N$>5$ per spectral pixel in the 3880~-~4020 \AA\ region; 10 galaxies are below this value and are removed from the sample (although one of these galaxies, 7549, has a strong Ca~II~K absorption which is clearly visible despite the low S/N).
    \item We excluded seven galaxies where Ca~II~H is completely hidden by H$\epsilon$ emission, preventing a robust measurement of the absorption line doublet properties (despite the detection of Ca~II~K absorption). In at least one case (galaxy 11494) the emission appears to be due to an imperfect fit to the stellar continuum rather than to actual H$\epsilon$ emission. We verify that the exclusion of galaxies with strong H$\epsilon$ does not change the average sSFR of the sample in an appreciable way.
\end{itemize}
The final sample consists of nine galaxies in which excess Na~I absorption due to neutral gas is present by selection. We also detect excess absorption in Ca~II in every galaxy, as discussed in Section~\ref{sec:fits}.
The distribution of sSFR and stellar mass for the selected galaxies is shown in Fig.~\ref{fig:sample} (red points), and the values are listed in Table~\ref{table_gal_values}. 
Our sample is located in the highest mass region of the survey and includes both star-forming and quiescent systems. The mass range covered is $10.6~ \lesssim~\log(M_{*}/M_{\odot})~ \lesssim~11.7$ and the redshift range is $1.8~\lesssim~z~\lesssim~2.8$. A comparison to the parent sample of 30 galaxies with neutral gas detected in Na~I by \citet[][blue squares in the figure]{Davies_2024} shows that the main reason for the lack of low-mass galaxies in our sample is that Na~I absorption by neutral gas is not detected in these systems. We also investigate the sample of Blue Jay galaxies without Na~I detection, finding only two tentative detections of excess absorption in Ca~II~H,~K, but the S/N is too low to perform a robust measurement of the doublet properties.

Fig.~\ref{fig:observed_spec_9395} shows a limited portion of the full spectral range for one galaxy, COSMOS 9395. The spectrum includes the Ca~II H, K and Na~I D absorption lines analyzed in this study, together with other important emission and absorption lines such as the Balmer series, [O~III]$\lambda\lambda$4960, 5008$\AA$ and [N~II]$\lambda\lambda$6549, 6585$\AA$. The observed NIRspec spectrum (blue curve) and the best-fit stellar continuum provided by \texttt{Prospector} (dark pink curve) are not perfectly matched at the Ca~II and Na~I wavelengths. To verify this quantitatively, for each galaxy we calculate the Ca~II~K equivalent width in the observed spectrum and then in 100 stellar models randomly drawn from the posterior. The observed equivalent width is always more than 1~\AA\ larger than the stellar model value, while the variation in equivalent width among the models drawn from the posterior is on average 0.6~\AA.
Hence, the stellar continuum model is not able to completely reproduce these absorption lines, and a neutral gas component is needed to explain the observed absorption excess.

\begin{figure*}[t]
    \centering
    \includegraphics[width=0.85\linewidth]{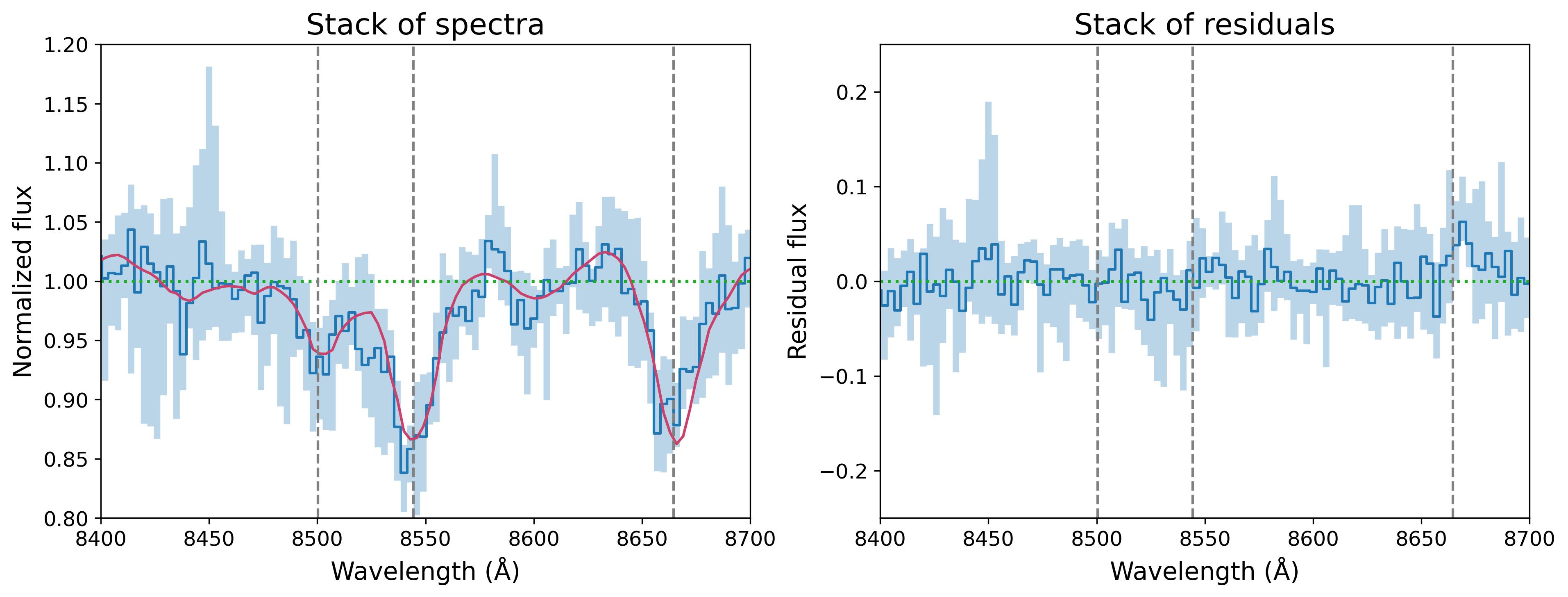}
    \caption{Stack of Ca~II triplet lines for the sample galaxies. \textit{Left panel}: Stack of the observed spectra (blue) and stack of the \texttt{Prospector} stellar models (dark pink) for the nine galaxies in the sample, around the Ca~II triplet. The stack is normalized so that the level of the continuum is approximately 1. The Ca~II triplet is precisely reproduced by the stellar models. \textit{Right panel}: Stack of the residuals, i.e., observed spectra divided by their best-fit stellar models. The dashed vertical lines mark the wavelengths of the absorption lines that are part of the Ca~II triplet. Shaded areas represent the 1$\sigma$ uncertainty.}
    \label{fig:stack}
\end{figure*}

\subsection{Validating the stellar model}

The measurements of the neutral gas properties are entirely dependent on the accuracy of the stellar model derived with \texttt{Prospector}, since we adopt that model to normalize the observed NIRSpec data. In principle, a biased stellar continuum model would lead to spurious residuals in the normalized spectrum, which we would interpret as real absorption by neutral gas. This could happen, for example, because of variations in the [Ca/Fe] abundance in high-redshift galaxies, which cannot be captured by our best-fit stellar model because the FSPS library adopts a fixed solar abundance pattern. 
To test this possibility, we consider the Ca~II triplet absorption lines observed at 8500 $\AA$, 8544 $\AA$ and 8664 $\AA$. Since this triplet is not resonant, it can only originate in stellar photospheres and not in cold neutral gas. Any residuals in the spectrum around these wavelengths must therefore be due to an imprecise stellar model.

We detect at least one absorption line of the Ca~II triplet in each galaxy, but the lines are faint, on the order of $5-15\%$ of the continuum, and appear rather noisy. We thus stack the spectra of the nine galaxies in the 8400 $\AA$~-~8700 $\AA$ wavelength region, and show the result in the left panel of Fig.~\ref{fig:stack}.
The dark pink curve is the stack of the \texttt{Prospector} fits, and correctly reproduces the observed absorption lines. The right panel shows the residuals, i.e., the ratio of the observed stack to the stack of the \texttt{Prospector} fits. The stacked residual flux is roughly constant around zero, proving that the \texttt{Prospector} models of the stellar spectra are correct. This result is similar to that obtained by \citet{Davies_2024}, who performed an analogous test using the Mg~b absorption line in the Blue Jay subsample with Na~I detections.

The fact that the best-fit spectrum is able to correctly reproduce the observed Ca~II triplet represents an important validation of the method, and confirms that the stellar population models are not systematically biased due to, for example, non-solar [Ca/Fe] abundances. We thus conclude that any absorption seen in resonant lines in excess to the best-fit stellar model must be attributed to neutral gas.

\begin{figure*}[htbp] 
\centering
\begin{minipage}[c]{0.6\textwidth} 
\centering 
\includegraphics[width=\linewidth]{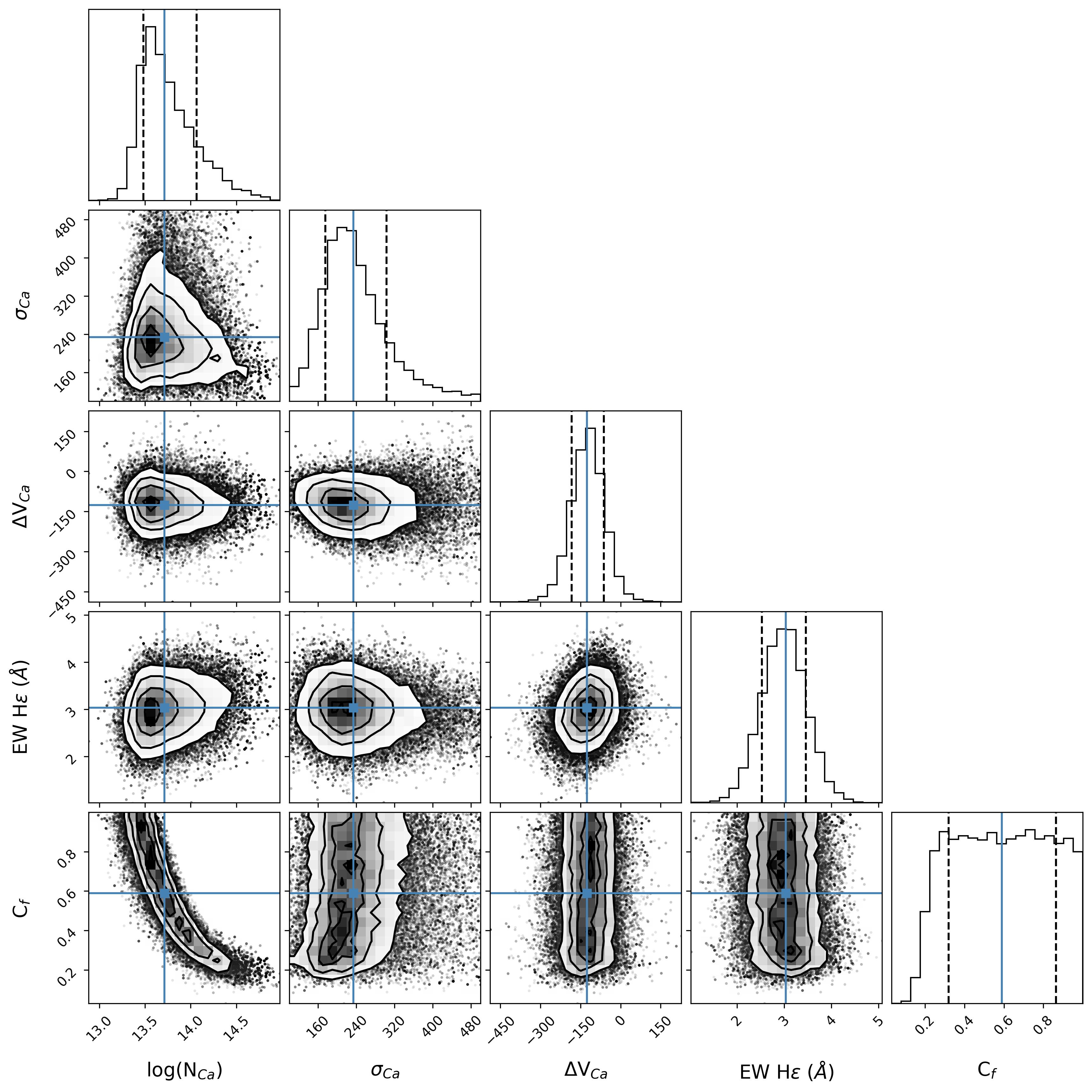} \end{minipage}
\hfill
\begin{minipage}[c]{0.35\textwidth} 
\caption{Corner plot showing the posterior distributions of the Ca~II parameters inferred by \texttt{emcee}, for galaxy COSMOS 18071. For the posterior distribution function of each parameter, the blue line is the median and the two dashed black lines are the 16$^{th}$ and the 84$^{th}$ percentiles.} \label{fig:corner_plot} 
\end{minipage} 
\end{figure*}

\section{Absorption line fitting} \label{abs_line_fitting}

\subsection{The model}

Since the absorption due to neutral gas constitutes a multiplicative term in the description of the observed galaxy spectrum, our first step is to divide each NIRSpec spectrum by the best-fit stellar continuum model obtained with \texttt{Prospector}. We then model, in the \say{normalized} spectrum, the excess absorption in Ca~II~H,~K and Na~I~D due to neutral gas, following the method outlined by \citet{Davies_2024} for their analysis of Na~I~D.

We used the \citet{Rupke_2005_sample} model:
\begin{equation}\label{eq_absline}
    I(\lambda) = 1 - C_{f} + C_{f} \cdot e^{-\tau_{\lambda}} \;,
\end{equation}
which describes the intensity of an absorption line assuming a unity continuum level. C$_{f}$ is the covering fraction along the line of sight, which we assume is constant with wavelength (i.e., with gas velocity). 
For the optical depth of a single absorption line, $\tau_{\lambda}$, we consider a Gaussian function with a central value given by
\begin{equation}
        \tau_{0} = 0.7580 \cdot \Bigg(\frac{N_{abs}}{10^{13} \ cm^{-2}}\Bigg)\Bigg(\frac{f_{lu}}{0.4164}\Bigg)\Bigg(\frac{\lambda_{lu}}{1215.7 \ \AA}\Bigg)\Bigg(\frac{10 \ km/s}{b}\Bigg),
    \label{tau0}
\end{equation}
where the oscillator strength $f_{lu}$ and the central rest-frame wavelength $\lambda_{lu}$ are fixed for each transition \citep{Draine_2011}. 
The column density of the absorbing element, $N_{abs}$, and the Doppler parameter $b = \sqrt{2} \sigma_{abs}$ (where $\sigma_{abs}$ is the absorption line velocity dispersion), are free to vary. We also introduced a free velocity shift $\Delta V_{abs}$ to account for the possible presence of outflows.

We develop a model that describes separately the Ca~II and Na~I doublets based on the Eq.~\ref{eq_absline} and \ref{tau0}. Each doublet consists of two absorption lines with different oscillator strength and rest-frame wavelength, but identical column density, velocity dispersion, and velocity shift relative to the galaxy systemic velocity. We also account for the presence of nearby emission lines, which contaminate the spectral region of interest (H$\epsilon~\lambda3970 \AA$ near Ca~II~H, and He~I$~\lambda 5875 \AA$ near Na~I); we thus introduced the following additive term to the model:
\begin{equation} 
    f(\lambda) = \frac{F_{em}}{\sqrt{2 \pi}\sigma_{em}} \cdot exp\Bigg(-\frac{(\lambda - \lambda_{em})^{2}}{2\sigma_{em}^{2}}\Bigg) \;.
\end{equation}
The shape of the emission line is fitted by a single Gaussian profile where $\sigma_{em}$ is the velocity dispersion and $F_{em}$ is the line flux. Since we are working with the normalized spectrum, in our model the flux $F_{em}$ coincides with EW$_{em}$, the emission line equivalent width in units of \AA. 
The emission line velocity is fixed to the galaxy systemic value, while the velocity dispersion $\sigma_{em}$ is fixed to the value measured from other ionized emission lines in the same spectrum (see Bugiani et al., in prep.), accounting for the nominal wavelength-dependent spectral resolution provided by the JWST documentation.

In principle, resonant transitions such as the Ca~II and Na~I doublets can produce both absorption and emission lines. The emission is typically redshifted, as it is produced by the receding side of the outflow \citep[e.g.,][]{Chen_2010, Concas_2019, Baron_2020}. We do not include this component in the model because we do not find evidence for redshifted emission in our spectra. It is possible that weak resonant emission is present and is partially infilling the absorption. This would cause a slight underestimate of the absorption line equivalent width and of the mass of the neutral outflow.

In conclusion, the model describing an absorption line doublet and its neighboring emission line is characterized by five free parameters: logarithmic column density of absorbing material log$_{10}$(N$_{abs}$); velocity shift and dispersion of the absorption lines $\Delta V_{abs}$ and $\sigma_{abs}$; covering fraction C$_{f}$; intensity of the contaminant emission line EW$_{em}$.

\begin{figure*}[t]
        \includegraphics[scale=0.35]{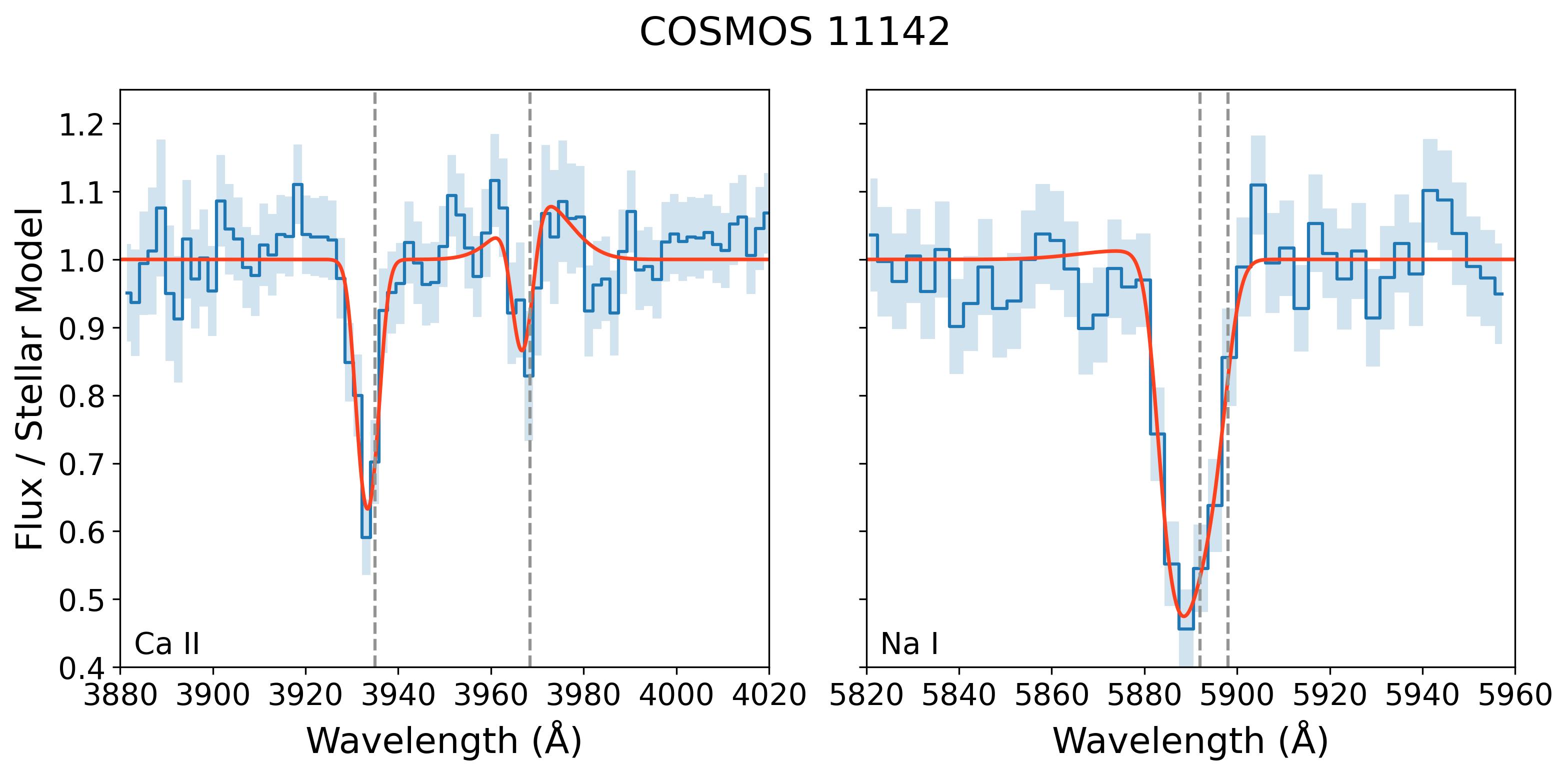}
        \includegraphics[scale=0.35]{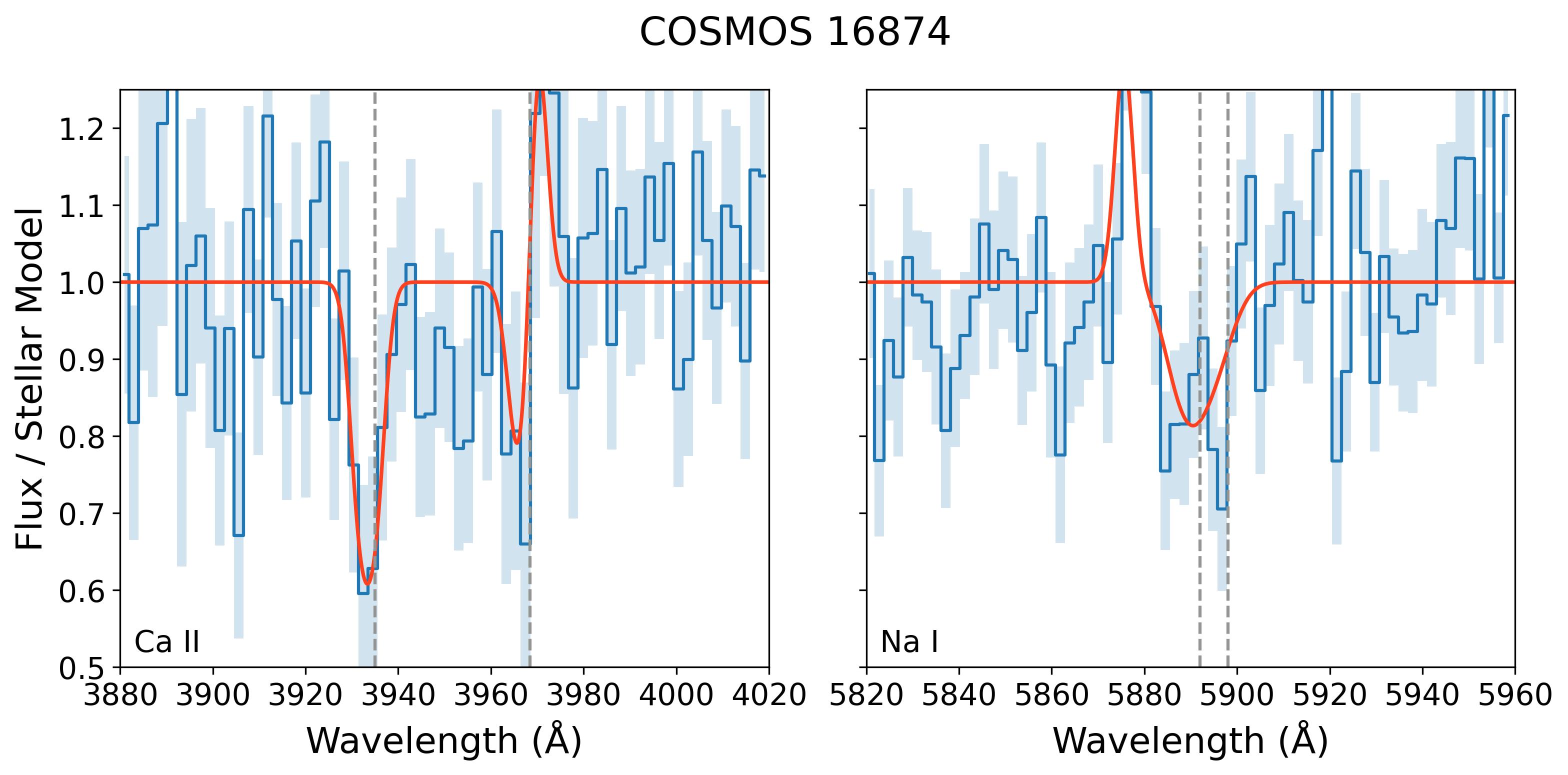}
    \caption{Observed NIRSpec spectra divided by their stellar continuum (blue), with the best-fit model (orange) for galaxies 11142 and 16874. The blue shadow is the flux uncertainty. For each galaxy the left panel shows the Ca~II~K,~H absorption and H$\epsilon$ emission, while the right panel shows the Na~I~D absorption and He~I emission. The dashed vertical lines mark the systemic wavelength of the absorption lines.}
    \label{fig:gal_spectra_2}
\end{figure*}

\subsection{Fitting results}
\label{sec:fits}

For each galaxy, we fit our model separately for Ca~II~H,~K (with H$\epsilon$ in emission) in the range 3880~-~4020~\AA, and for Na~I~D (with He~I in emission) in the range 5820~-~5960~\AA.
We perform a Markov chain Monte Carlo (MCMC) analysis \citep{mcmc,mcmc_rev} with the \texttt{emcee} code \citep{Foreman_Mackey_2013}, and infer the best-fit value for the free parameters of the model.
We adopted Jeffrey's prior for the column density and the velocity dispersion, and uniform priors for the other parameters, with the following ranges:
\begin{itemize}
    \item[$\bullet$] log$_{10}$(N$_{abs}$/cm$^{-2}$): [11, 15]; 
    \item[$\bullet$] $\sigma_{abs}$: [100, 500] km/s;
    \item[$\bullet$] $\Delta V_{abs}$: [-500, 300] km/s;
    \item[$\bullet$] C$_{f}$: [0, 1];
    \item[$\bullet$] EW$_{em}$: [0, 10] \AA. 
\end{itemize}

The covering fraction and the column density are notoriously degenerate; but this degeneracy can be broken by the relative depth of the two absorption lines in a doublet. Thus, we leave C$_f$ free when fitting the Ca~II doublet, in which the two lines are well resolved. The Na~I doublet is, however, unresolved, and for this reason we fix C$_f$ for Na~I to the value measured for Ca~II in the same galaxy.
This assumption is justified by the fact that the two doublets are tracing the same neutral gas (as discussed in Section~\ref{kinematics}).

Fig.~\ref{fig:corner_plot} shows an example of the posterior distribution for the model parameters obtained from fitting Ca~II and H$\epsilon$ in the normalized spectrum of one galaxy in our sample. The degeneracy between the covering fraction and the column density is clearly visible in the bottom left panel. The degeneracy is such that the product of these two parameters is approximately constant along the posterior distribution. In this case, the fit is unable to strongly constrain the covering fraction despite having a well-resolved Ca~II doublet, because the H$\epsilon$ emission line is relatively strong and can \say{hide} an arbitrary amount of absorption from the underlying Ca~II~H line.

We show the Ca~II and Na~I doublets, together with the best-fit from our model, for two representative galaxies in Fig.\ref{fig:gal_spectra_2} (spectra for all sample galaxies are illustrated in Appendix~\ref{appendix_A}). The best-fit model shown in the figure includes both the absorption lines due to neutral gas and the contaminating emission lines due to ionized gas. Looking at Fig.\ref{fig:gal_spectra_2}, it is clear that the Ca~II doublet is well resolved, unlike the Na~I doublet. However, the H$\epsilon$ emission line, which is almost exactly on top of the Ca~II~H absorption line, in some cases adds substantial uncertainty to the fit results (as in right panel of Fig.\ref{fig:gal_spectra_2}). The Na~I doublet is contaminated by He~I emission, but this is less severe because He~I is slightly offset in wavelength compared to the absorption lines.

\begin{figure*}
    \centering
    \includegraphics[width=0.43\linewidth]{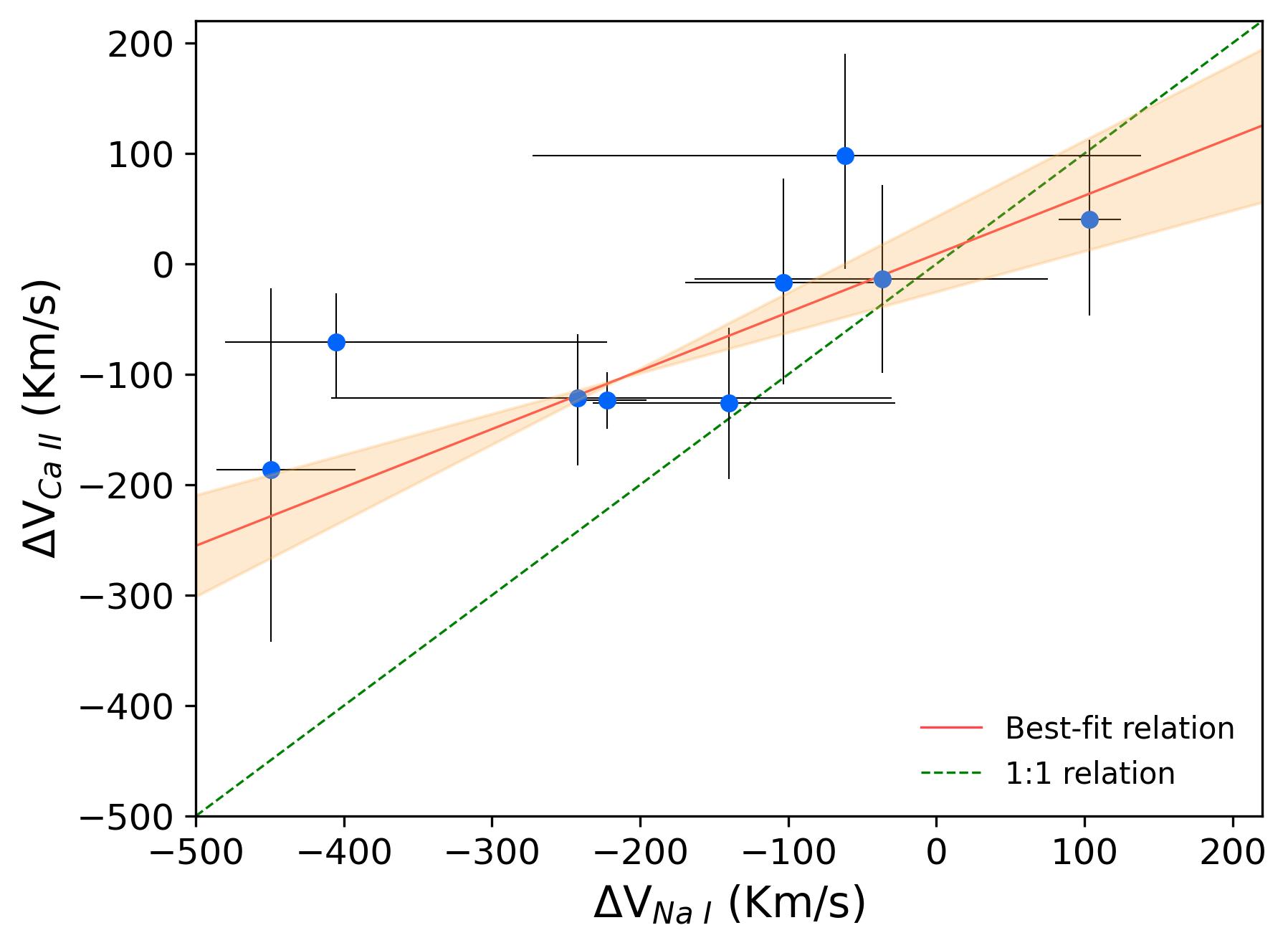}
    \includegraphics[width=0.43\linewidth]{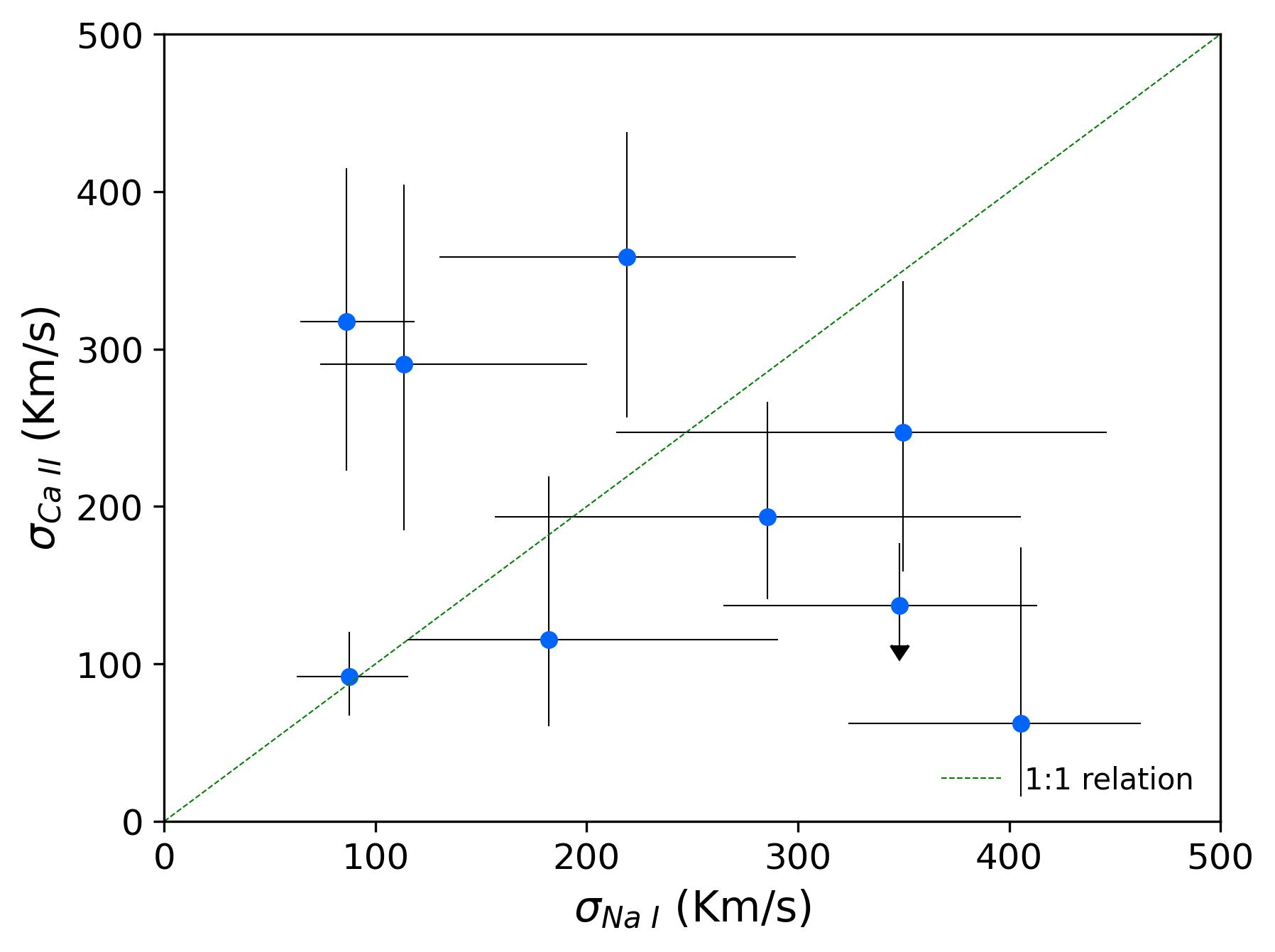}
    \caption{Kinematic properties of neutral gas. \textit{Left panel}: Velocity shifts for neutral gas measured from Ca~II vs. Na~I. Error bars correspond to the 16$^{th}$ and the 84$^{th}$ percentile of the probability distribution function from \texttt{emcee}. The red line represents the best-fit linear relation of the measurements weighted by their errors, while the orange shadow is the 1$\sigma$ error of the best-fit line. \textit{Right panel}: Velocity dispersions of neutral gas inferred from Ca~II vs. Na~I. The dashed green line represents the 1:1 relation. The velocity dispersion of Ca~II lines of COSMOS-18668 is only an upper limit represented by an arrow.}
    \label{fig:kinematics}
\end{figure*}
The absorption line parameters derived for Na~I and Ca~II are listed in Table~\ref{table_gal_values} for our sample. The velocity dispersions in the table have been corrected for the nominal instrumental resolution, which in some cases may be underestimated because it was derived for uniform slit illumination \citep{Degraaff_2024_resolution}.
The kinematics and column densities measured for Na~I are broadly consistent with those obtained by \citet{Davies_2024} from the same data, despite the use of a slightly different methodology, because in this work we first derive the covering fraction from Ca~II and then apply it to Na~I. We also note that the best-fit covering fractions are generally high, $C_f \sim 0.4 - 0.7$, but with large uncertainties due to degeneracy resulting in a broad range of possible values, $C_f \sim 0.2 - 0.9$.

\section{Kinematics of neutral gas} \label{kinematics}

The neutral gas detected via excess absorption in the Ca~II~H,~K lines may be part of the galaxy ISM, or may be found in an outflow.
In order to discriminate between these two possibilities, we need to analyze the gas kinematics: outflows observed in absorption are always in front of the galaxy and therefore are blueshifted with respect to the systemic velocity measured from the stellar spectrum. We find that five out of nine galaxies have a blueshifted absorption from the Ca~II fit; three are consistent with the systemic velocity; only 1 is redshifted. A similar breakdown is obtained when analyzing the Na~I fit results. This is a first indication that the two doublets trace the same neutral gas, which in many cases is part of an outflow.

We compare the velocity shifts measured from Ca~II and Na~I, $\Delta V_{Ca~II}$ and $\Delta V_{Na~I}$ respectively, in the left panel of Fig.~\ref{fig:kinematics}. There is a clear correlation between the two measurements. This is a strong indication that the Ca~II H, K absorption lines trace neutral gas that is in similar conditions to those of the gas traced by Na~I~D. Incidentally, this is also a further validation of our stellar population modeling -- if the excess absorption was due to an imperfect subtraction of the stellar component, we would not expect to see such a clear correlation in the Ca~II and Na~I kinematics.  
The red line in the figure represents the best-fit relation $\Delta V_{Ca \ II} = (0.5 \pm 0.2 )\cdot \Delta V_{Na \ I} \ + \ (9 \pm 34)$~km/s, which we obtained accounting for the uncertainties in both the $x$ and the $y$ axis. The detection of an outflow in the Ca~II lines is generally consistent with outflows detected in Na~I. Interestingly, the only galaxy with a redshifted Na~I absorption, COSMOS 19572, also has a redshifted Ca~II absorption, even though it is consistent with the systemic value when the uncertainty is taken into account. JWST imaging of this galaxy reveals that it is undergoing a merger, thus supporting the possibility of a neutral gas inflow \citep{Davies_2024}.

Based on the result obtained for the velocity shifts, we expect to also observe a correlation between the velocity dispersions measured from Ca~II and Na~I. However, as shown in the right panel of Fig.~\ref{fig:kinematics}, this correlation is not present, and the $\sigma_{Ca \ II}$ versus $\sigma_{Na \ I}$ measurements appear to be randomly distributed.
This is a result of the large uncertainty in the measured velocity dispersion, which is on the order of 50\%, fully consistent with the observed scatter in the $\sigma_{Ca \ II}$ versus $\sigma_{Na \ I}$ relation.
It is plausible that such large uncertainties may also systematically affect the velocity shift measurement, although to a lesser extent, and thus explain why the relation between $\Delta V_{Ca \ II}$ and $\Delta V_{Na \ I}$ deviates from the 1:1 relation. 
As a test, we repeated the absorption line fits fixing the velocity dispersion of Na~I~D to that measured from Ca~II~H and K, which should be more robust because of the wider wavelength separation between the doublet lines. However, this did not substantially change the relation between $\Delta V_{Ca \ II}$ and $\Delta V_{Na \ I}$. 
New NIRSpec observations with the high-resolution grating will be able to resolve the Na~I~D doublet, measure its kinematics to a higher degree of precision, and shed light on these issues. 

\begin{figure}[t]
    \centering
    \includegraphics[width=0.9\linewidth]{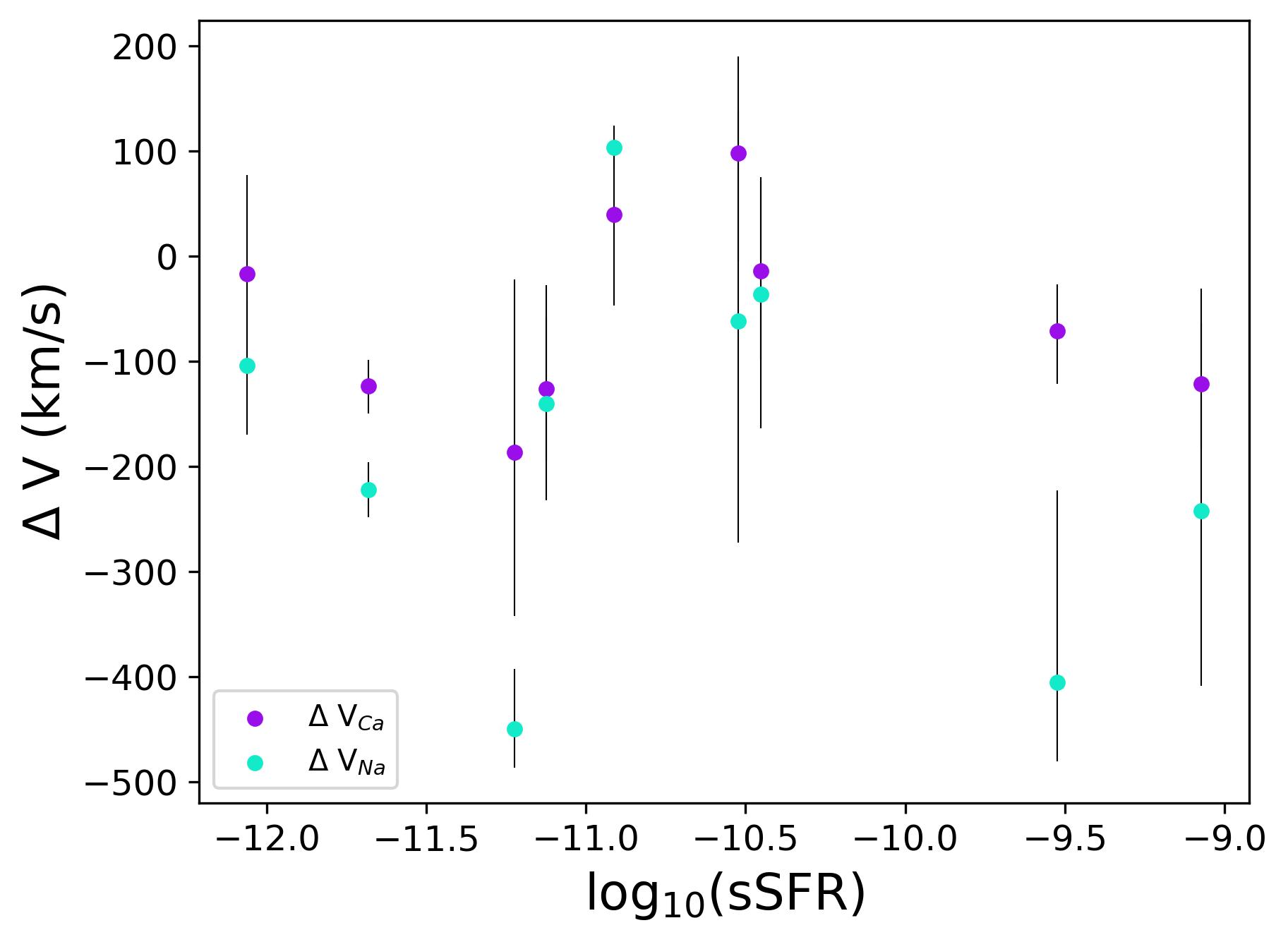}
    \caption{Velocity shift vs sSFR, shown for both the Ca~II and Na~I lines. No trend is apparent, suggesting the stellar feedback is likely not responsible for the bulk of the observed neutral outflows.}
    \label{fig:sSFR_deltaV}
\end{figure}
Finally, we compare the gas velocity to the sSFR, looking for a possible correlation. As illustrated in Fig.\ref{fig:sSFR_deltaV}, we do not find a trend, suggesting that most of the outflows observed in our sample are due to AGN feedback and not stellar feedback. This is consistent with the results of \citet{Davies_2024} based on the ionized emission line properties in the larger Blue Jay sample, from which our galaxies are drawn.

\begin{figure}[t]
    \centering
    \includegraphics[width=0.9\linewidth]{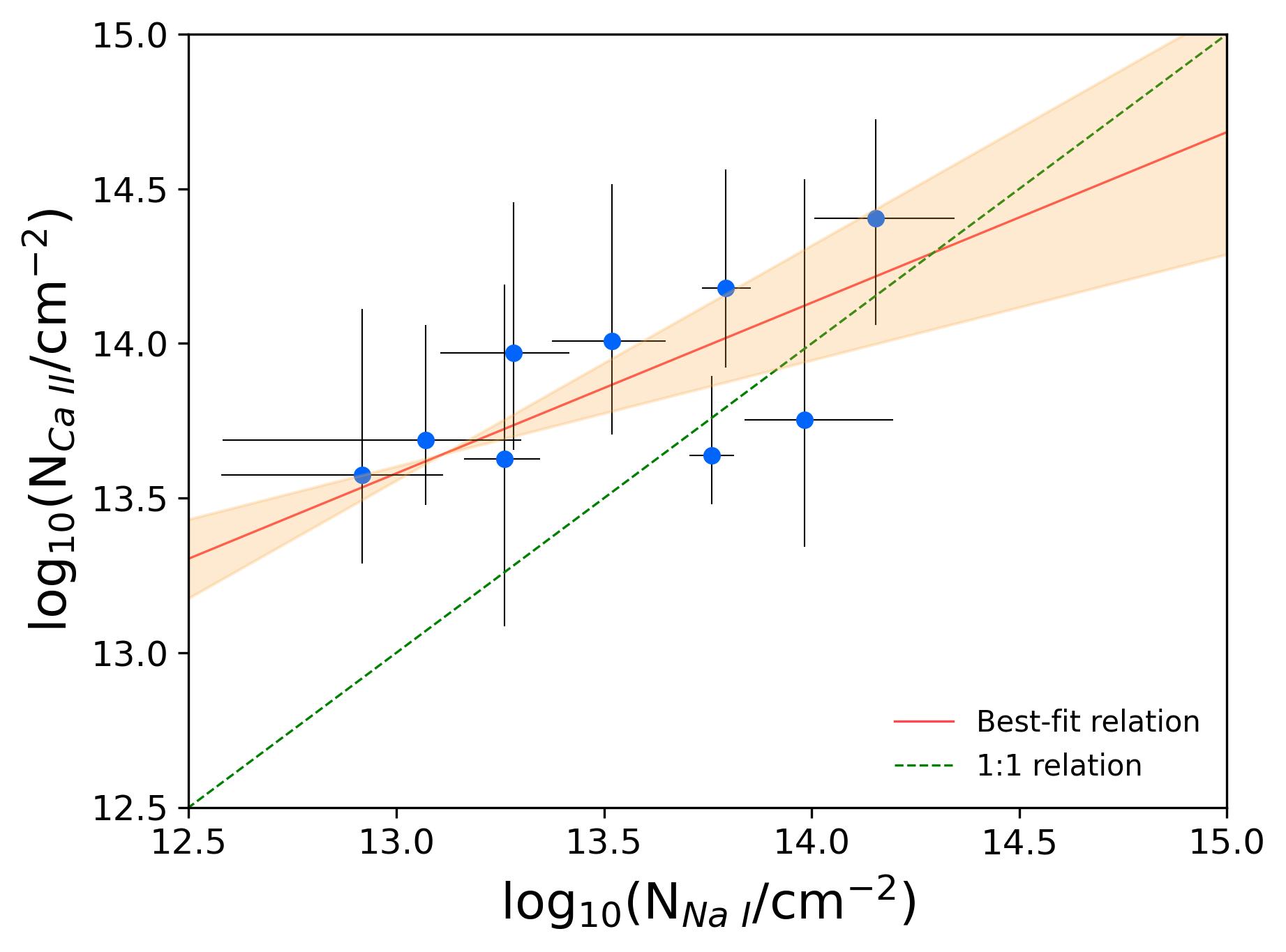}
    \caption{Relation between the column densities of Ca~II and Na~I. The red line represents the best-fit linear relation of the measurements weighted by their uncertainties. Error bars represent the 16$^{th}$ and 84$^{th}$ percentile of the posterior distribution.}
    \label{fig:col_dens}
\end{figure}

\section{Neutral gas column density}
\label{column_dens_section}

We compare the column densities of Ca~II and Na~I in Fig.~\ref{fig:col_dens}. A clear trend is present, which we fit with the following relation:
\begin{equation} \label{eq:col_dens}
    \log \NCa = (0.55 \pm 0.21) \cdot \log \NNa + (6.5 \pm 0.1),
\end{equation}
where the column densities are expressed in units of cm$^{-2}$.
This result is an additional confirmation that Ca~II~H,~K and Na~I~D trace similar types of gas. However, the relation is not 1:1, meaning that the relative proportion of Ca~II and Na~I atoms changes with the total amount of gas along the line of sight.

To further explore the relation between Ca~II and Na~I column densities, let us consider the relation between each of these metals and the column density of neutral hydrogen, \NH, which dominates the mass budget of neutral gas.
Following \citet{Rupke_2005_sample} we can write the Na~I column density as:
\begin{equation}
\label{eq:Nna}
    \NNa = \NH \cdot (1-y_{Na}) \; 10^{[Na/H]} \;(n_{Na}/n_{H})_{\odot} \; B_{Na} \; ,
\end{equation}
where $(1-y_{Na})$ is the ionization correction, [Na/H] is the Na abundance in the galaxy relative to the solar value, $(n_{Na}/n_{H})_{\odot}$ is the Na abundance in the Sun, and $B_{Na}$ is the dust depletion. Since most of the Na atoms are singly ionized, the ionization correction is substantial, $(1-y_{Na}) \approx 0.1$ \citep{Rupke_2005_sample}. In the case of Ca~II the ionization correction is negligible since we are directly probing the dominant ionization stage \citep[see, e.g.,][]{Murray_2007}, so we can write
\begin{equation}
\label{eq:Nca}
    \NCa \simeq \NH \cdot 10^{[Ca/H]} \;(n_{Ca}/n_{H})_{\odot} \; B_{Ca} \; .
\end{equation}
Given that the solar abundances of Ca and Na are very similar \citep{Asplund_2021, Magg_2022}, we can write the ratio of the column densities of the two elements as
\begin{equation}
\label{eq:ratio}
    \frac{\NNa}{\NCa} \simeq (1-y_{Na}) \; 10^{[Na/H]-[Ca/H]} \; \frac{B_{Na}}{B_{Ca}} \; .
\end{equation}
This ratio varies systematically with the gas column density because the relation between \NNa\ and \NCa\ is not 1:1. Unfortunately, the three factors in Eq.~\ref{eq:ratio} are extremely difficult to measure directly, and show a wide variation in local studies. The Na ionization correction is rarely measured and can vary by at least a factor of two \citep{Baron_2020}, while the Na abundance in local galaxies spans 0.5 dex \citep{Conroy_2014}. The dust depletion is even more uncertain and, for Ca, can vary by up to 4 dex depending on the amount of dust, as revealed by studies of Milky Way gas clouds and galaxies at $z < 0.5$  \citep{Hobbs_1974,Phillips_1984,Savage_1996,Guber_2016}.

\begin{figure}[t]
    \centering
    \includegraphics[width=0.9\linewidth]{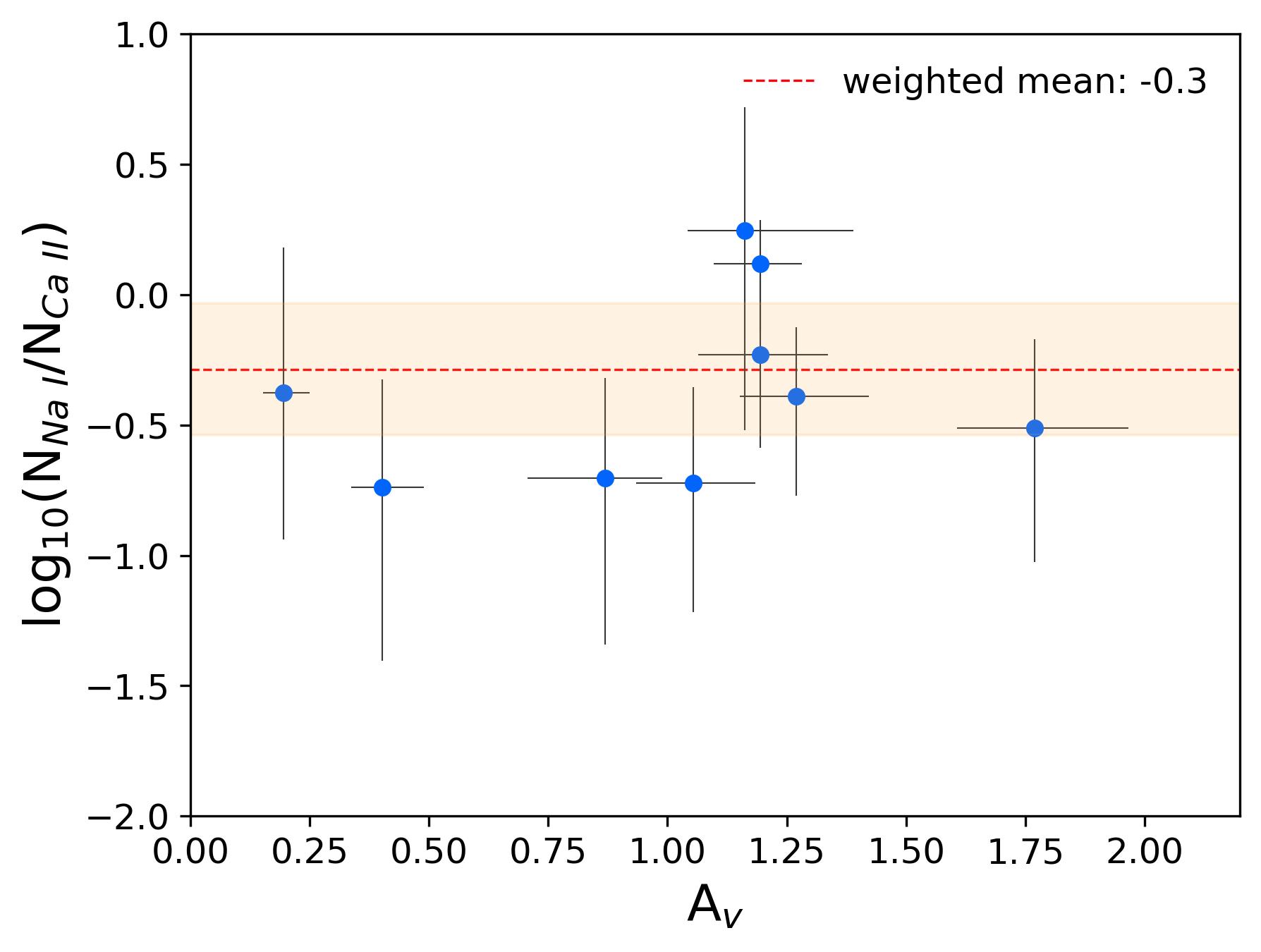}
    \caption{Logarithmic ratio of Ca and Na column densities as a function of dust attenuation A$_{V}$. The dashed red line marks the weighted mean, while the orange shadow is the error on the weighted mean.}
    \label{fig:dust_attenuation}
\end{figure} 

In order to explore the role of dust depletion, we investigate whether the Na-to-Ca column density ratio depends on the galaxy dust attenuation inferred by \texttt{Prospector}. This is shown in Fig.~\ref{fig:dust_attenuation}, where $A_V$ is the diffuse component of the dust in the ISM and is one of the two components in the model of \citet{Charlot_2000} employed by \texttt{Prospector} (we do not use the birth cloud component since the outflows are likely not co-spatial with the star forming regions). There is no systematic trend with dust attenuation, and galaxies are scattered around the value of $-0.3$ dex, which is the weighted mean of the column density ratio. This suggests that the variation of the Na-to-Ca column density ratio is not simply due to dust depletion; however, we note that the dust attenuation measured for a galaxy by \texttt{Prospector} does not necessarily reflect the amount of dust present in the neutral gas outflow.   
Similar results are obtained for other physical parameters measured with \texttt{Prospector}: we do not detect trends between the column density ratio and the SFR, the mass, or the age of the galaxy.

\begin{figure}[t]
    \centering
    \includegraphics[width=0.9\linewidth]{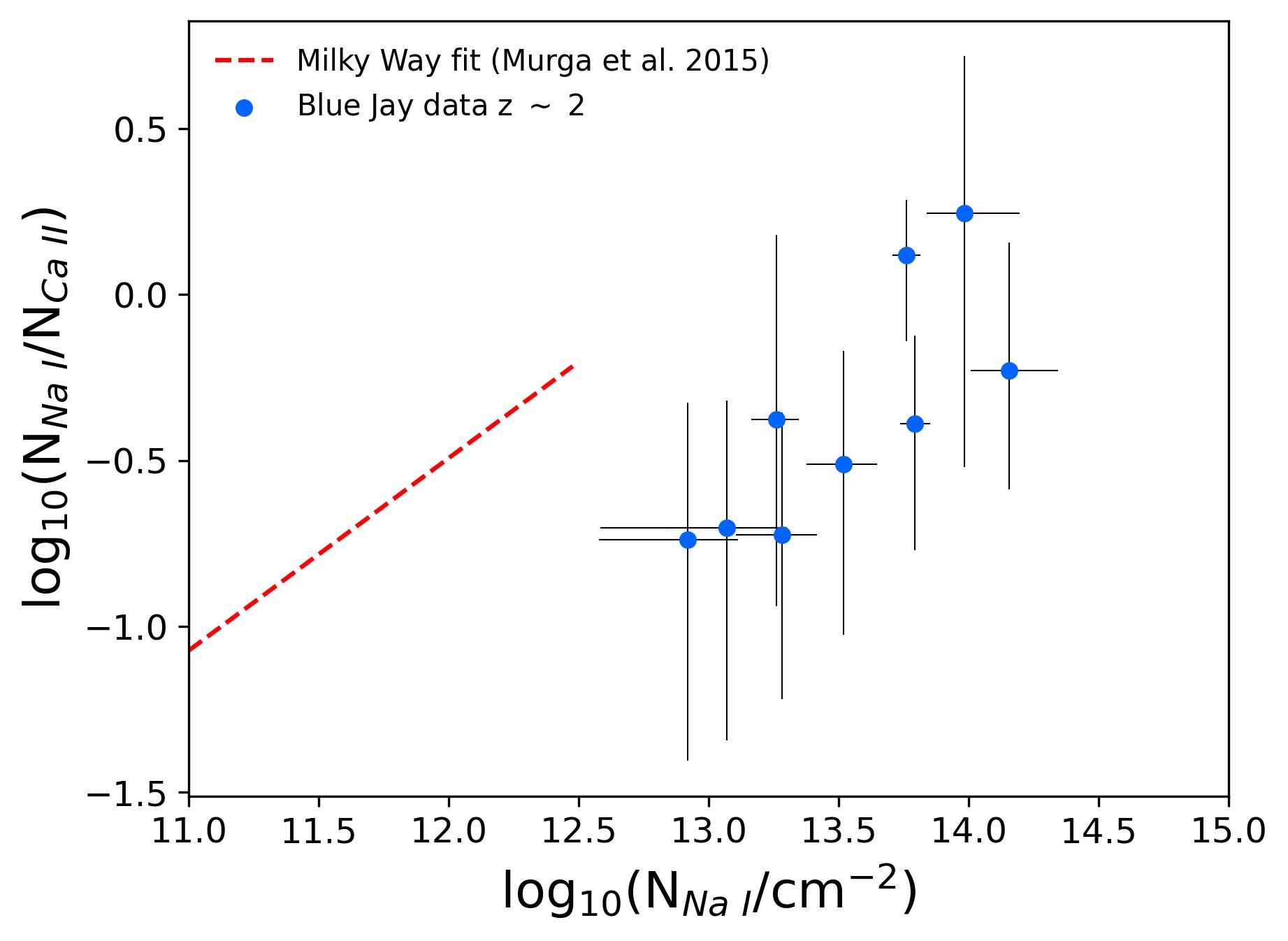}
    \caption{Column density ratio vs.~\NNa. The dashed red line represent the correlation between these quantities empirically derived for Milky Way clouds \citep{Murga_2015}.}
    \label{fig:MW_relation}
\end{figure}

A statistically robust relation between the Ca~II and Na~I column density has been derived for neutral gas clouds in the Milky Way by \citet{Murga_2015}: 
\begin{equation}\label{eq:Murga}
    \NNa/\NCa = \Bigg(\frac{\NNa}{N_{2}}\Bigg)^{\alpha_{2}},
\end{equation}
with $N_{2} = (7.07 \pm 0.82) \cdot 10^{12}$ cm$^{-2}$ and $\alpha_{2} = 0.58 \pm 0.03$. We compare this relation with the results for our sample of $z\sim2$ galaxies in Fig.~\ref{fig:MW_relation}.
Our galaxies follow a trend with a similar slope ($0.7\pm0.3$) to the \citet{Murga_2015} relation, but with a large offset: at high redshift, galaxies lie at a lower Na-to-Ca ratio compared to an extrapolation of the local relation.
This is not surprising because we are comparing Milky Way measurements performed on very small spatial scales to high-redshift measurements taken over the entire galaxy extent.
For example, an overestimate of the covering fraction would move our points to lower values of \NNa\ while leaving the column density ratio unchanged, thus alleviating the tension with the local relation. However, we can robustly exclude that the covering fractions are overestimated by an order of magnitude because the maximum depth reached by the neutral gas absorption lines gives a hard lower limit of $C_f \sim 0.2 - 0.4$, which is not much lower than the best-fit values.

One possibility is that the discrepancy observed in Fig.~\ref{fig:MW_relation} simply reflects the different physical conditions found in $z\sim2$ galaxies compared to Milky Way clouds.
However, another possibility is that the observed neutral gas in high-redshift galaxies is clumpy: in this case the larger column density observed at high redshift is due to a larger number of clumps, and not to a change in their physical conditions. This would shift the points towards larger \NNa\ values without altering the column density ratio.

\section{Tracing neutral outflows with Ca~II~H,~K}
\label{discussion}

Given that both the velocity and the column density of Ca~II are tightly correlated to those of Na~I, we conclude that the Ca~II~H and K lines can be used to trace neutral gas outflows in alternative to, or together with, the widely used Na~I~D doublet.
The main challenge of estimating the total amount of gas, which is mostly in the form of H atoms, from observations of a trace element remains. This requires some form of conversion between \NCa\ and \NH. Using a theoretical conversion based on Eq.~\ref{eq:Nca} would lead to systematic uncertainties of orders of magnitude given the poorly constrained dust depletion of Ca. Instead, we use our observations to derive a fully empirical conversion between the Ca~II and H~I column densities.

We start with our best-fit relation between Ca~II and Na~I column density, given in Eq.~\ref{eq:col_dens}, which has a relatively small scatter of 0.24 dex. We invert this relation and obtain an inferred column density of Na~I, given Ca~II~H, K observations. Next, we need to convert this to a column density of hydrogen; one way to achieve this is by taking the theoretical relation (Eq.~\ref{eq:Nna}) and adopting nominal assumptions about ionization correction, abundance, and dust depletion. Instead, we adopt the empirical conversion derived by \citet{Moretti_2025} by direct measurement of H~I and Na~I column density for the neutral outflow in a $z=2.45$ galaxy, which was made uniquely possible by the alignment with a background quasar:
\begin{equation}
    \label{eq:moretti}
    \log \NH = \log \NNa + 7.5 \,.
\end{equation}
We thus obtained a fully empirical conversion between the Ca~II and H~I column density:
\begin{equation}
\label{eq:calibration}
    \log \NH \approx 1.8 \cdot \log \NCa - 4.3 \; .
\end{equation}
This relation can be used to estimate the H~I column density in high-redshift galaxies when only the Ca~II absorption lines are available.
We expect this relation to hold for all massive galaxies ($\log M_{*}/M_{\odot} > 10.6 $) at $z\sim2$, because the Blue Jay sample is representative by design, and in this work we further selected galaxies mostly based on stellar mass and without strong biases in star formation rate or other galaxy properties. Naturally, the relation given in Eq.~\ref{eq:calibration} can only be trusted within the range of column densities probed by our sample, which is $13.5 < \log (\NCa/\mathrm{cm}^{-2}) < 14.5$.

\subsection{Hydrogen column density} \label{sec:N_H}

We now infer \NH\ through the application of three distinct methods, and using both Na~I and Ca~II absorption lines. The methods used for the calculation are the following: 
\begin{enumerate}
    \item Empirical relations: Both Eq.~\ref{eq:moretti} and Eq.~\ref{eq:calibration} are based on the high-redshift Na~I~D calibration by \citet{Moretti_2025};
    \item Fixed dust depletion: We adopted the commonly used equations Eq.~\ref{eq:Nna} and Eq.~\ref{eq:Nca} by \citet{Rupke_2005_sample}, considering a constant dust depletion $\delta_{Na} = -0.95$ and $\delta_{Ca} = -2.04$ (the latter being the mean dust depletion derived by \citet{Phillips_1984});
    \item \citet{Wakker_2000}: This work relates the elemental abundances to the hydrogen column density \NH\ for the interstellar medium in the Milky Way:
\begin{equation}\label{eq:wakker_Na}
    \log(\NH) \approx 1.2\, \log(\NNa) - 5.6.
\end{equation}
\begin{equation}\label{eq:wakker_Ca}
    \log(\NH) \approx 4.5\, \log(\NCa) - 38.9;
\end{equation}
\end{enumerate}

\begin{figure}
    \centering
    \includegraphics[width=0.9\linewidth]{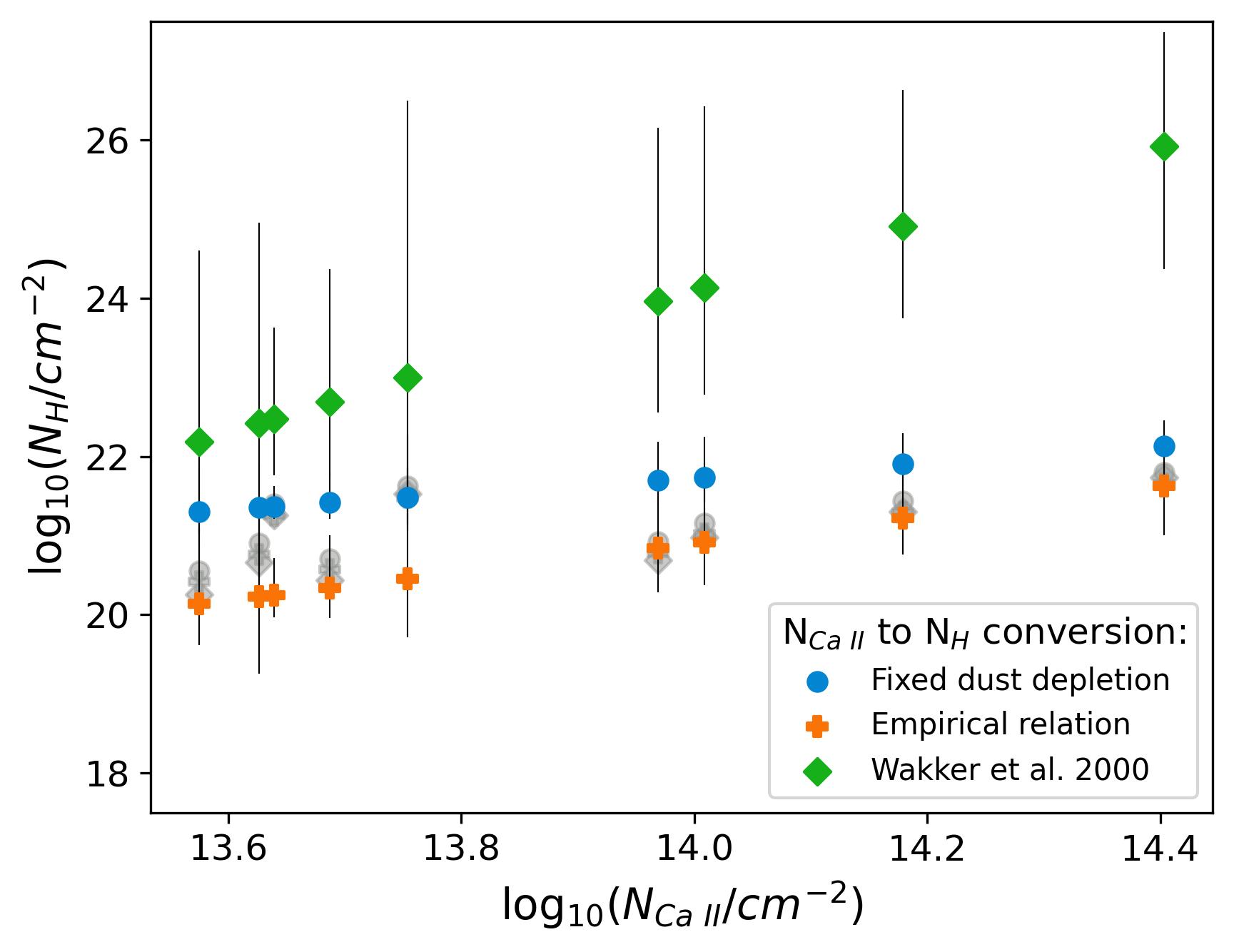}
    \caption{Comparison between log(\NH) inferred by different equations as a function of log(\NCa). The three methods, and the corresponding equations, are described in Section~\ref{sec:N_H}. Gray points on the background are the measure of log(\NH) based on Na~I absorption lines for each equation.}
    \label{fig:log_Ca_Vs_log_Nh}
\end{figure}

Table~\ref{table_outflow_properties} lists the hydrogen column densities estimated with the different methods and using different absorption lines. The results obtained from Na~I using different methods are highly consistent, and are in the range $\log(\NH) \approx 20.4-21.8$. The results obtained with Ca~II, instead, show substantial variations, with \NH\ ranging varying by more 2-4 dex according to which method is adopted, as illustrated in Fig.~\ref{fig:log_Ca_Vs_log_Nh}.
The estimate of the hydrogen column density computed from Eq.\ref{eq:Nca} strongly depends on the dust depletion, which is highly uncertain in the case of Ca. Moreover, as the value of \NCa\ increases, the discrepancy between Eq.~\ref{eq:wakker_Ca} and the others grows due to the extrapolation of a Milky Way study to high-redshift galaxies. This suggests that the more robust method is that based on direct calibration at $z \sim 2 $, i.e., Eq.~\ref{eq:moretti} and Eq.~\ref{eq:calibration}.
Fig.~\ref{fig:log_Ca_Vs_log_Nh} also shows that \NH\ values derived from the application of the empirical equation for Ca~II match within the uncertainties the results obtained from Na~I (shown in gray), by construction. 
We thus adopt the empirical relation for estimating \NH\ from the observed absorption lines in the remainder of this work.

\begin{figure}
    \centering
    \includegraphics[width=0.8\linewidth]{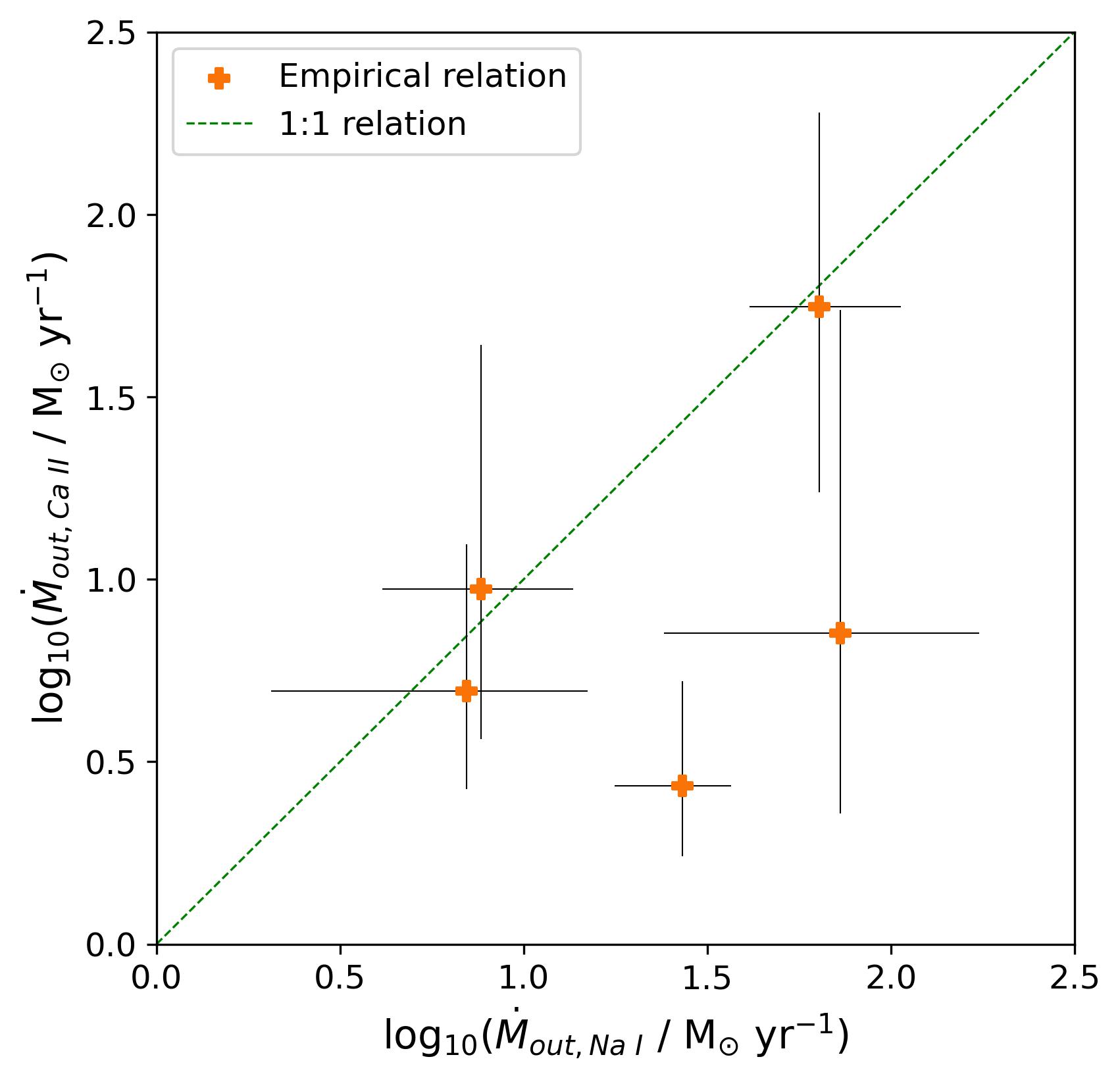}
    \caption{Mass outflow rates derived by Ca~II absorption lines vs. Mass outflow rate derived by Na~I absorption lines in logarithmic scale. The computation is performed only for \NH\ values inferred by Eq.~\ref{eq:moretti} and Eq.~\ref{eq:calibration} (the empirical relations). The 1:1 relation is shown as a dashed green line.}
    \label{fig:mass_out_rate}
\end{figure}

\subsection{Outflow mass and mass outflow rate from Ca II} \label{mass_rate}

Once the H~I column density is known, the outflow mass and the mass outflow rate for each galaxy exhibiting a velocity shift compatible with the presence of an outflow are calculated through the following equations \citep{Rupke_2005_analysis}:
\begin{equation}\label{eq:mass_and_rate}
\begin{split}
    M_{out} = & 1.4 m_{p} \cdot 4\pi \, C_\Omega \, C_f \, \NH \, R_{out}^{2} \; ;\\
    \dot{M}_{out} = & 1.4 m_{p} \cdot 4\pi \, C_\Omega \, C_f \, \NH \, R_{out}v_{out} \; ,
\end{split}
\end{equation} 
where $R_{out}$ and $v_{out}$ are the radius and velocity of the outflow, and $C_\Omega$ represents the fraction of solid angle covered by the outflow. The radial extension of outflows at $z \geq 1$ can vary between $\lesssim 1$~kpc and $\sim3$~kpc \citep{Cresci_2023,Veilleux_2023,deugenio_2024}.
\begin{table*}
\renewcommand{\arraystretch}{1.7}
\caption{Hydrogen column densities and mass outflow rates derived from Na~I and Ca~II.}
\label{table_outflow_properties}
\centering
\begin{tabular}{ccccccccc}
\hline \hline
\ & \multicolumn{2}{c}{Fixed dust depletion} & \multicolumn{2}{c}{\citet{Wakker_2000}} & \multicolumn{2}{c}{Empirical relation} & \multicolumn{2}{c}{Empirical relation}\\
\hline
IDs & logN$_{H}$ & logN$_{H}$ & logN$_{H}$ & logN$_{H}$ & logN$_{H}$ & logN$_{H}$& $\dot{M}_{out}$ & $\dot{M}_{out}$ \\
\hline
\ & from Na~I & from Ca~II & from Na~I & from Ca~II & from Na~I & from Ca~II & from Na~I & from Ca~II \\
\hline
 \ & Eq.~\ref{eq:Nna} & Eq.~\ref{eq:Nca} & Eq.~\ref{eq:wakker_Na} & Eq.~\ref{eq:wakker_Ca} & Eq.~\ref{eq:moretti} & Eq.~\ref{eq:calibration} & M$_{\odot}$/yr & M$_{\odot}$/yr\\
\hline
8002 & 21.6$^{+0.2}_{-0.1}$ & 21.5$^{+0.8}_{-0.4}$ & 21.5$^{+0.3}_{-0.2}$ & 23.0$^{+3.5}_{-1.8}$ & 21.5$^{+0.2}_{-0.1}$ & 20.5$^{+1.4}_{-0.7}$ & 73.0$^{+100.9}_{-48.9}$ & 7.1$^{+47.7}_{-4.8}$ \\
18668 & 21.8$^{+0.2}_{-0.1}$ & 22.1$^{+0.3}_{-0.3}$ & 21.7$^{+0.2}_{-0.2}$ & 25.9$^{+1.4}_{-1.5}$ & 21.7$^{+0.2}_{-0.1}$ & 21.6$^{+0.6}_{-0.6}$ & 63.9$^{+42.7}_{-22.6}$ & 55.9$^{+134.0}_{-38.5}$ \\
11142 & 21.4$^{+0.1}_{-0.1}$ & 21.4$^{+0.3}_{-0.2}$ & 21.3$^{+0.1}_{-0.1}$ & 22.5$^{+1.2}_{-0.7}$ & 21.3$^{+0.1}_{-0.1}$ & 20.3$^{+0.5}_{-0.3}$ & 27.0$^{+9.8}_{-9.3}$ & 2.7$^{+2.5}_{-1.0}$ \\
16874 & 20.9$^{+0.1}_{-0.2}$ & 21.7$^{+0.5}_{-0.3}$ & 20.7$^{+0.2}_{-0.2}$ & 24.0$^{+2.2}_{-1.4}$ & 20.8$^{+0.1}_{-0.2}$ & 20.8$^{+0.9}_{-0.6}$ & 7.7$^{+6.0}_{-3.6}$ & 9.4$^{+34.6}_{-5.7}$ \\
9395 & 20.6$^{+0.2}_{-0.3}$ & 21.3$^{+0.5}_{-0.3}$ & 20.3$^{+0.2}_{-0.4}$ & 22.2$^{+2.4}_{-1.3}$ & 20.4$^{+0.2}_{-0.3}$ & 20.1$^{+1.0}_{-0.5}$ & - & - \\
18071 & 20.7$^{+0.2}_{-0.5}$ & 21.4$^{+0.4}_{-0.2}$ & 20.4$^{+0.3}_{-0.6}$ & 22.7$^{+1.7}_{-0.9}$ & 20.6$^{+0.2}_{-0.5}$ & 20.3$^{+0.7}_{-0.4}$ & 7.0$^{+7.9}_{-4.9}$ & 4.9$^{+7.5}_{-2.3}$\\
16419 & 20.9$^{+0.1}_{-0.1}$ & 21.4$^{+0.6}_{-0.5}$ & 20.7$^{+0.1}_{-0.1}$ & 22.4$^{+2.5}_{-2.4}$ & 20.8$^{+0.1}_{-0.1}$ & 20.2$^{+1.0}_{-1.0}$ & - & - \\
10245 & 21.2$^{+0.1}_{-0.1}$ & 21.7$^{+0.5}_{-0.3}$ & 21.0$^{+0.2}_{-0.2}$ & 24.1$^{+2.3}_{-1.4}$ & 21.0$^{+0.1}_{-0.1}$ & 20.9$^{+0.9}_{-0.5}$ & - & - \\
19572 & 21.4$^{+0.1}_{-0.1}$ & 21.9$^{+0.4}_{-0.3}$ & 21.3$^{+0.1}_{-0.1}$ & 24.9$^{+1.7}_{-1.2}$ & 21.3$^{+0.1}_{-0.1}$ & 21.2$^{+0.7}_{-0.5}$ & - & - \\
\hline
\end{tabular}
\tablefoot{The hydrogen column densities are computed through several equations illustrated in Sec.~\ref{sec:N_H}. The errors are 1$\sigma$ of the measure.}
\end{table*}
Due to the inability to directly observe the radial extension, we assume $R_{out} = 1$~kpc, yielding a potential underestimation of the mass outflow rates. We also set $C_\Omega=0.5$, according to neutral outflows studies in local infrared galaxies \citep{Rupke_2005_sample}. Notably, the column density enters these equations only as the product $C_f \NH$, which substantially reduces the impact of the degeneracy with the covering fraction, since the absorption line observations are able to constrain this product much more precisely compared to $C_f$ and \NH\ individually, as discussed in Sec.~\ref{sec:fits}. To be consistent with results from \citet{Davies_2024} and facilitate comparison, the outflow velocity for our sample of galaxies is defined as $v_{out} = |\Delta V| + 2\sigma$, where $\Delta V$ is the velocity shift of the centroid and $\sigma$ the velocity dispersion of the absorption line.

The mass outflow rates are reported in Table~\ref{table_outflow_properties} and are shown in Fig.~\ref{fig:mass_out_rate}. The ones measured from Na~I, with a median value of $\dot{M}_{out, Na~I}\approx 39.3$, are in agreement with the results reported by \citet{Davies_2024} for a larger sample of galaxies from the Blue Jay survey, and consistent with properties of neutral outflows measured in star-forming galaxies and AGN host galaxies both in local Universe and at Cosmic Noon \citep{Rupke_2005_sample,Baron_2020, Roberts_Borsani_2020,Cresci_2023}. Nevertheless, the outflow rates inferred from Ca~II, with a median value of $\dot{M}_{out,Ca~II}\approx 7.6$, show a discrepancy of a factor of $\sim5$ with the ones derived by Na~I. This could be caused by the different ionization level, and thus temperature range, traced by the two lines \citep[e.g.,][]{Valentino_2025}. However, the uncertainties on these measurements are large, and may at least partially explain the observed discrepancy.

Comparing the mass outflow rates from Na~I and Ca~II with the SFRs of the five galaxies with an outflow, we notice that for four out of five galaxies the mass outflow rate is higher than its corresponding SFR. Consequently, the presence of outflows in these galaxies could play an important role in the quenching of star formation.

\section{Summary and conclusions} \label{conlusions}

In this work we have analyzed deep JWST/NIRSpec spectra (R $\sim$ 1000) from the Blue Jay survey for nine quiescent and star-forming galaxies with $10.6 < \log(M_{*}/M_{\odot}) < 11.7$ and $1.8<z<2.8$. Our main objective is to probe the neutral gas, which in these galaxies has already been detected via Na~I~D absorption, using the Ca~II~H, K absorption line doublet. After removing the stellar contribution to the observed spectra we have fit a model of the neutral gas absorption to the wavelength regions around the Na~I~D and Ca~II~H, K lines, to measure the kinematics and column densities. Our main results are the following:
\begin{itemize}
    
    \item[$\bullet$] The velocity shifts measured from Ca~II and Na~I absorption lines are clearly correlated, indicating that the two elements trace gas in similar physical conditions. Neutral outflows traced by Na~I can therefore be studied also using Ca~II. The velocity dispersions are not correlated, but this is consistent with the large error bars in the measurements, partly due to the relatively low $R\sim1000$ spectral resolution which leads to a poorly resolved Na~I~D doublet.

    \item[$\bullet$] The column densities of Ca~II and Na~I are also correlated, supporting the idea that they trace similar gas phases. However, the relation is not 1:1, meaning that the \NNa/\NCa\ ratio varies systematically with column density. This may depend on the dust depletion of Ca atoms, but we do not find a trend between the column density ratio and the galaxy dust attenuation, nor on any other galaxy properties. Compared to the local relation derived for gas clouds in the Milky Way, our sample shows a similar slope but a systematic offset, which may suggest the presence of a clumpy medium with a large number of clouds.
    
    \item[$\bullet$] We make use of our observed relation between the Ca~II and Na~I column density, together with a recently published direct measurement of the Na-to-H column density ratio, to derive an empirical conversion between the Ca~II and the H~I column density, given in Eq.~\ref{eq:calibration}. This calibration is used to estimate the properties of neutral gas outflows, such as outflow mass and mass outflow rate, which are consistent within a factor of five with previous studies. 

\end{itemize}

This work is the first systematic investigation of neutral gas in high-redshift galaxies based on Ca~II~H,~K, and further studies will likely improve our understanding of the physical processes that set the Ca~II column density and its relation with other galaxy properties. 
New JWST/NIRSpec observations at a higher spectral resolution targeting Na~I~D in this sample (GO 5427; PI Davies) will soon enable a more accurate comparison of the Ca~II and Na~I properties, potentially shedding light on some of the open questions.

Our work offers a new way to estimate the properties of neutral outflows. The Ca~II doublet has the advantage of being easily resolved in medium-resolution spectra, but it can also be heavily contaminated by the H$\epsilon$ line.
Ca~II can be used in alternative to, or together with, other absorption lines such as Na~I~D and Mg~II. Having access to a wide range of spectral features to study the neutral phase is crucial in order to limit the systematic uncertainties and to expand as much as possible the sample of galaxies with at least one measurement of neutral gas. This gas phase appears to play a key role in the rapid quenching of massive galaxies \citep{Belli_2024, deugenio_2024, Wu_2025} and, as JWST observations reveal quiescent galaxies at increasingly higher redshift \citep{Carnall_2024,Degraaff_2024, Weibel_2025}, it becomes crucial to characterize neutral outflows in the earliest phases of cosmic history.

\begin{acknowledgements}
The Blue Jay survey is funded in part by STScI Grant JWST- GO-01810. 
SB, LB, AHK and MS are supported by ERC grant 101076080 \say{Red Cardinal}. 
RLD is supported by the Australian Research Council through the Discovery Early Career Researcher Award (DECRA) Fellowship DE240100136 funded by the Australian Government. 
RW acknowledges funding of a Leibniz Junior Research
Group (project number J131/2022).   
This research was supported by the International Space Science Institute (ISSI) in Bern, through ISSI International Team project 24-602 "Multiphase Outflows in Galaxies at Cosmic Noon".
This work is based on observations made with the NASA/ESA/ CSA James Webb Space Telescope. The data were obtained from the Mikulski Archive for Space Telescopes at the Space Telescope Science Institute, which is operated by the Association of Universities for Research in Astronomy, Inc., under NASA contract NAS 5-03127 for JWST. These observations are associated with program GO 1810. This work also makes use of observations taken by the 3D-HST Treasury Program (GO 12177 and 12328) with the NASA/ESA HST, which is operated by the Association of Universities for Research in Astronomy, Inc., under NASA contract NAS 5-26555.
\end{acknowledgements}

\bibliographystyle{aa} 
\bibliography{bibliography} 

\clearpage
\appendix
\section{Sample galaxy spectra}\label{appendix_A}
Spectra divided by the stellar continuum from \texttt{Prospector} of Ca~II ad Na~I doublets for all galaxies, with their best-fit from our model, are shown in Fig.\ref{fig:gal_spectra}. The best-fit, represented by the orange line, reproduces both absorption and emission lines in the selected spectral region. Some Ca~II and Na~I doublet spectra (e.g., galaxy 18071) are contaminated by H$\epsilon$ (left panels) and He~I (right panels) emission lines, which contribute to the uncertainties in the absorption line measurements.
\begin{center}
\begin{minipage}{\textwidth}
    \centering
    
    \includegraphics[width=\linewidth, height=4.5cm, keepaspectratio]{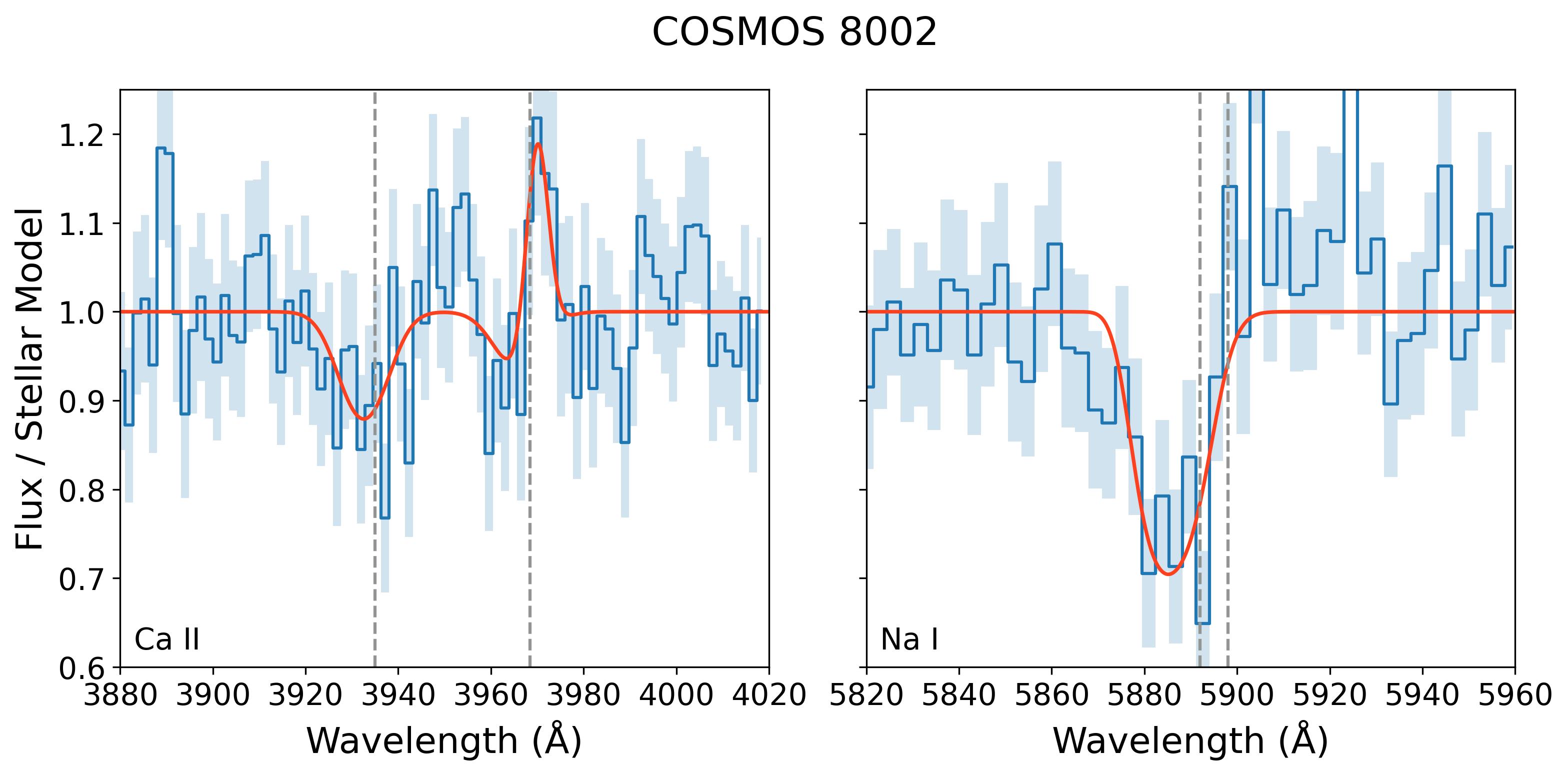}
    \includegraphics[width=\linewidth, height=4.5cm, keepaspectratio]{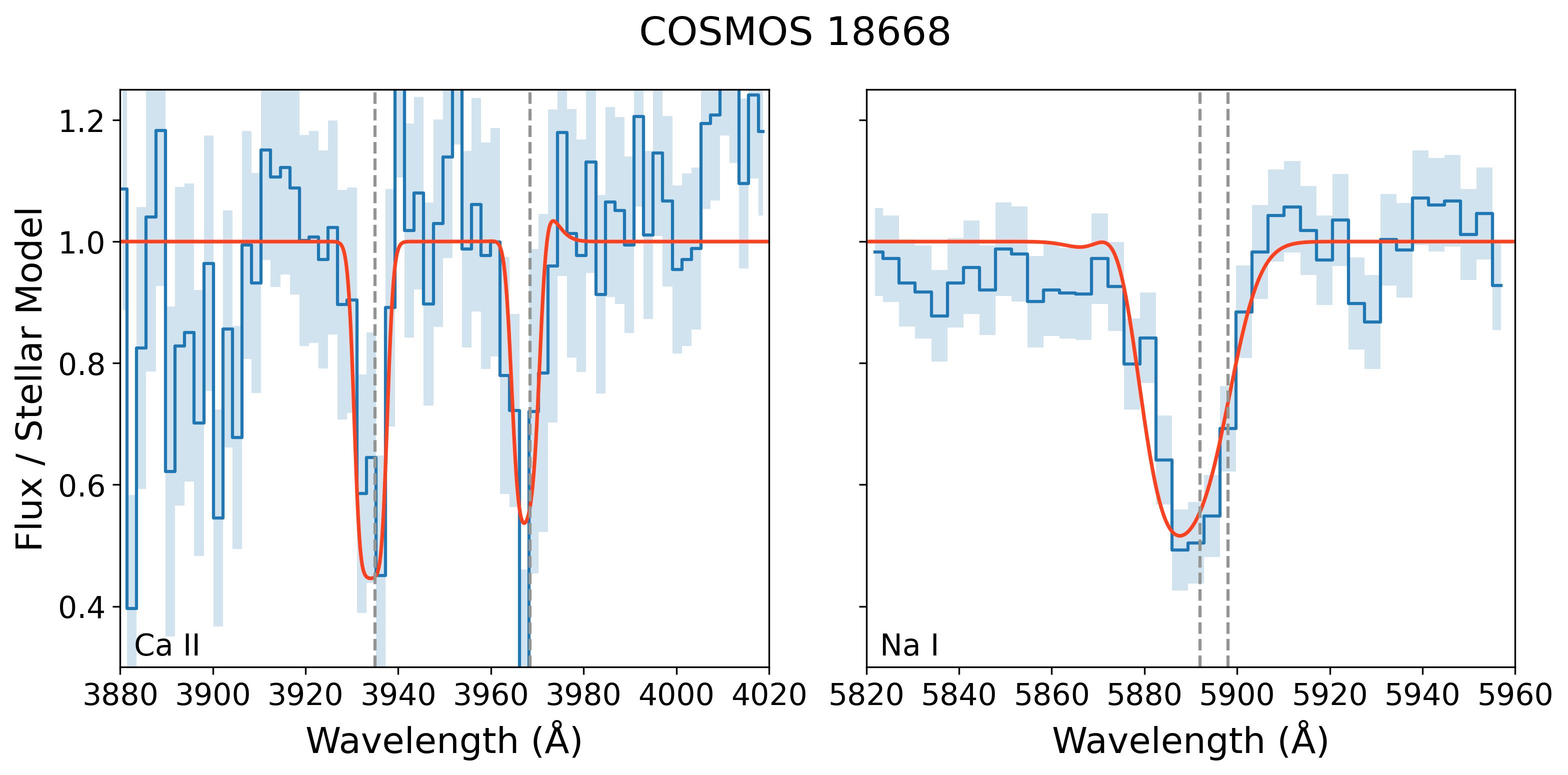}
    \includegraphics[width=\linewidth, height=4.5cm, keepaspectratio]{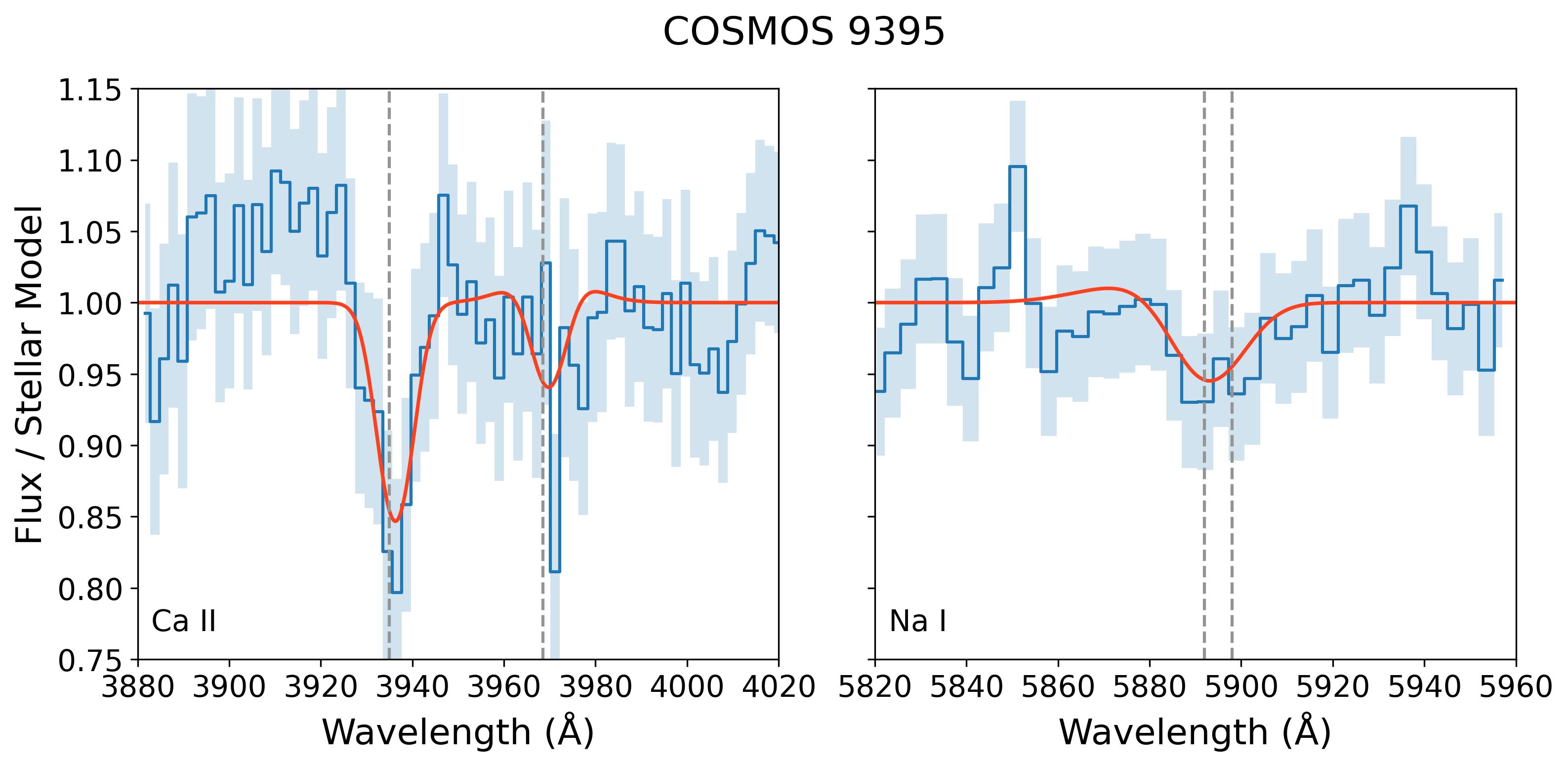}
    \includegraphics[width=\linewidth, height=4.5cm, keepaspectratio]{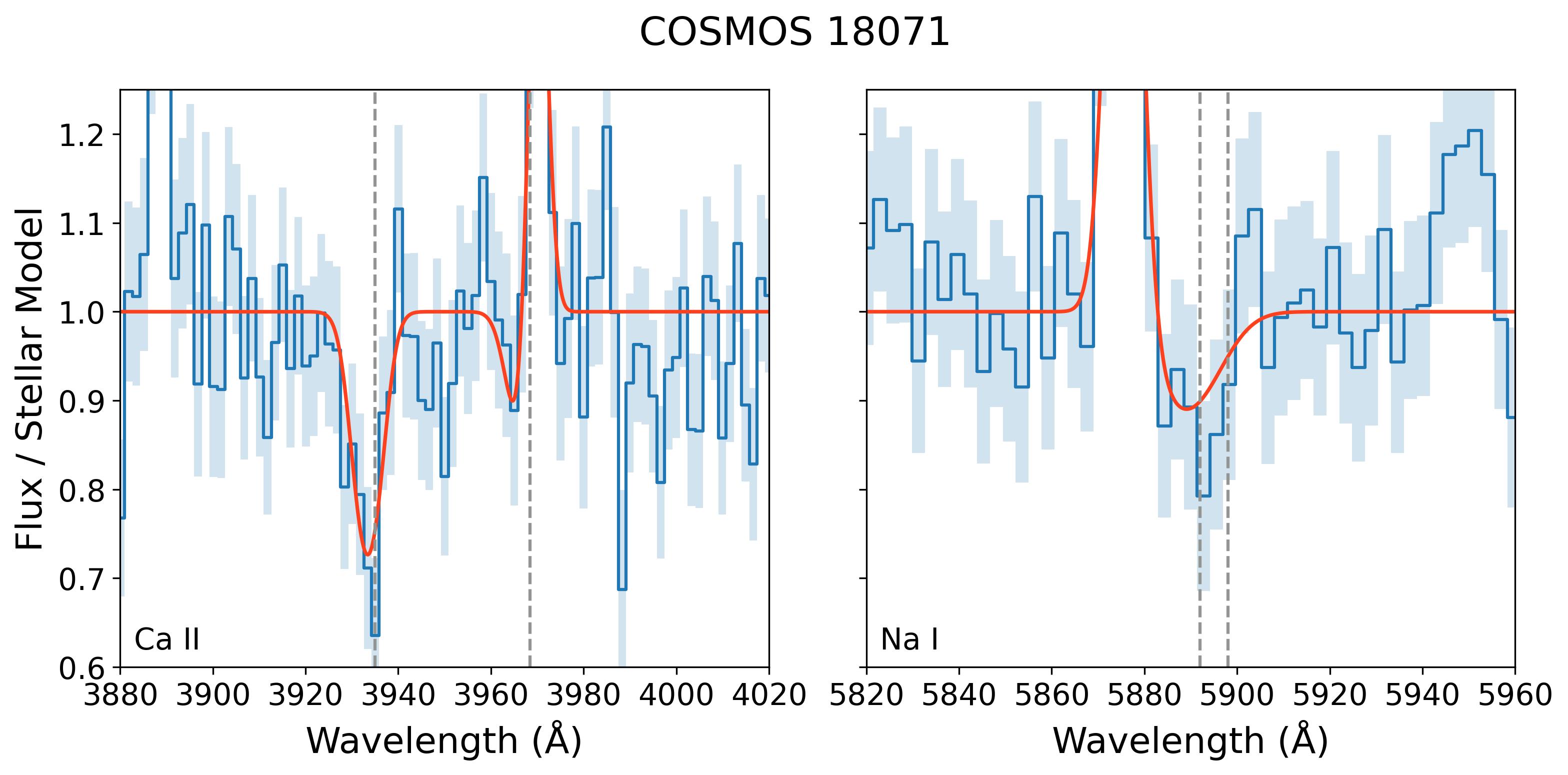}
    \includegraphics[width=\linewidth, height=4.5cm, keepaspectratio]{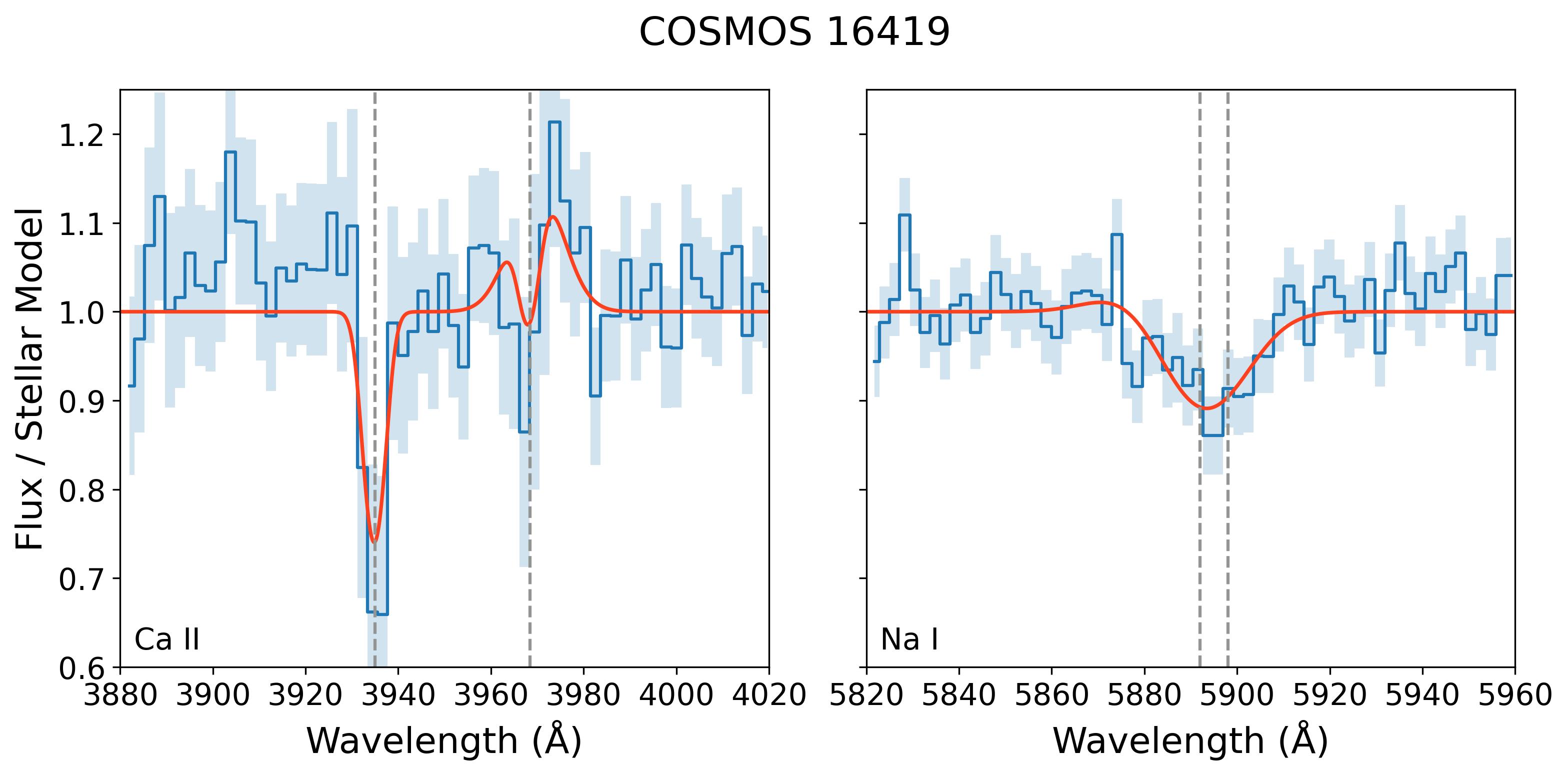}
    \includegraphics[width=\linewidth, height=4.5cm, keepaspectratio]{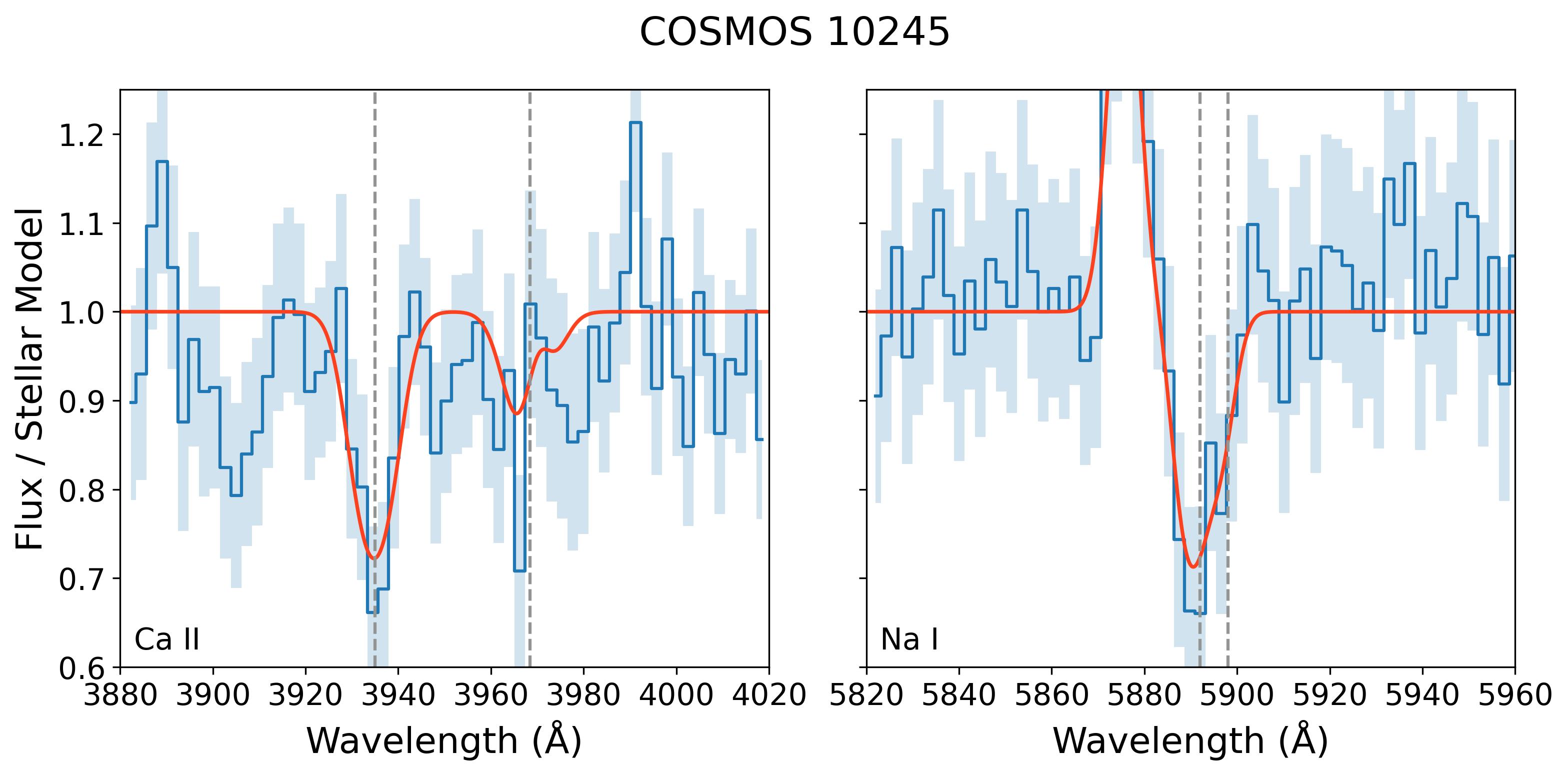}
    \includegraphics[width=\linewidth, height=4.5cm, keepaspectratio]{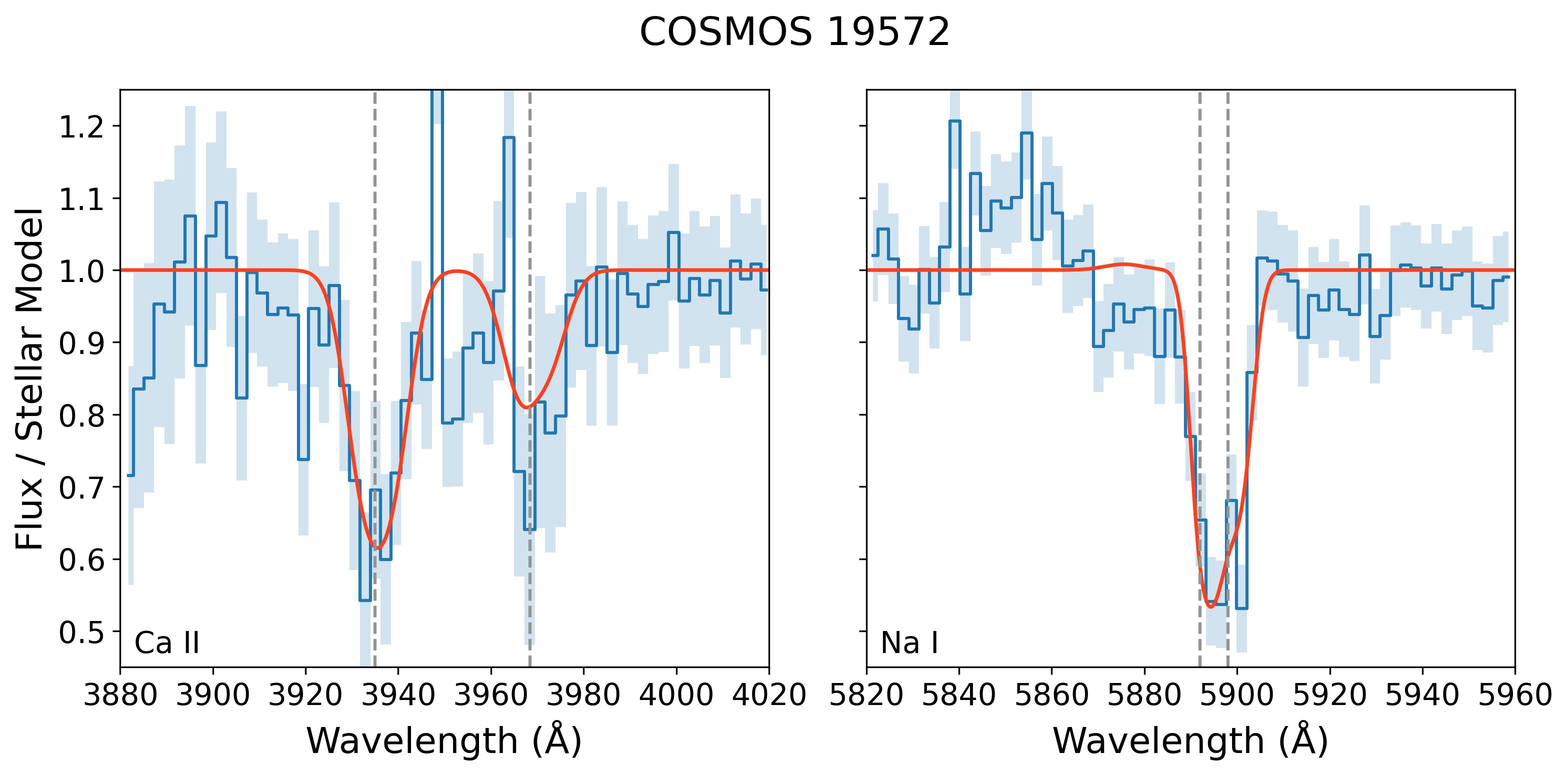}

    \captionof{figure}{Observed NIRSpec spectra of the other seven galaxies divided by their stellar continuum (blue), with the best-fit model (orange). The blue shadow is the flux uncertainty. For each galaxy the left panel shows the Ca~II~K,~H absorption and H$\epsilon$ emission, while the right panel shows the Na~I~D absorption and He~I emission. The dashed vertical lines mark the systemic wavelength of the absorption lines.}
    \label{fig:gal_spectra}
\end{minipage}
\end{center}

\end{document}